\PassOptionsToPackage{hyphens}{url}
\PassOptionsToPackage{usenames,dvipsnames,svgnames,table}{xcolor}
\documentclass[format=sigconf,letterpaper,screen=true]{acmart}
\usepackage[utf8]{inputenc}
\usepackage[unicode]{hyperref}
\usepackage{xspace}
\usepackage{xfrac} \usepackage{caption}
\usepackage{color}
\usepackage{colortbl}
\usepackage{graphicx}
\usepackage{booktabs}
\usepackage{multirow}
\usepackage{tumcolor}
\usepackage{tabularx}
\usepackage{balance} \usepackage[group-digits=true, group-separator={,}]{siunitx}
\usepackage[pointedenum]{paralist}
\usepackage{url}
\usepackage{listings}
\usepackage{cleveref}
\usepackage{comment}
\usepackage{subcaption} \usepackage{subfloat} \usepackage{pifont} \usepackage{algorithmic}
\usepackage{algorithm}

 \algsetup{linenodelimiter=}

\usepackage{microtype}
\microtypesetup{expansion=alltext,step=1} \tolerance=300         \pdfpagewidth=8.5in
\newif\ifstatus
\statustrue
\newcounter{nalg}[section] \renewcommand{\thenalg}{\thesection .\arabic{nalg}} \DeclareCaptionLabelFormat{algocaption}{Algorithm \thenalg}

\newcommand{\latinlocution}[1]{\textit{#1}}

\newcommand{\eg}{\latinlocution{e.g.},\xspace}
\newcommand{\ie}{\latinlocution{i.e.},\xspace}
\newcommand{\cf}{\latinlocution{cf.},\xspace}
\newcommand{\etal}{\latinlocution{et al.}\xspace}

\newcommand{\first}{\emph{(i)}\xspace}
\newcommand{\second}{\emph{(ii)}\xspace}

\newcommand{\one}{\emph{(i)}\xspace}
\newcommand{\two}{\emph{(ii)}\xspace}
\newcommand{\three}{\emph{(iii)}\xspace}

\newcommand{\xref}[1]{\hyperref[#1]{\S\ref*{#1}}\xspace}

\newcommand{\topm}{Top\,1M\xspace}
\newcommand{\topk}{Top\,1k\xspace}
\newcommand{\toph}{Top\,100\xspace}
\newcommand{\toplist}{top list\xspace}
\newcommand{\toplists}{top lists\xspace}

\colorlet{tablelight}{black!10}
\colorlet{tablemid}{black!30}
\colorlet{tabledark}{black!50}

\newcommand{\redtrid}{\textcolor{BrickRed}{\ding{116}}}

\newcommand{\greentriu}{\textcolor{Green}{\ding{115}}}

\newcommand{\bluensq}{\textcolor{Blue}{\ding{110}}}

\newcommand\para[1]{\textbf{#1:}}

\crefname{section}{\S}{\S}
\Crefname{section}{\S}{\S}
\crefformat{section}{\S#2#1#3}

\makeatletter
\newcommand\nocaption{    \renewcommand\p@subfigure{}
    \renewcommand\thesubfigure{\thefigure\alph{subfigure}}
}

\copyrightyear{2018} 
\acmYear{2018} 
\setcopyright{acmlicensed}
\acmConference[IMC '18]{2018 Internet Measurement Conference}{October 31-November 2, 2018}{Boston, MA, USA}
\acmBooktitle{2018 Internet Measurement Conference (IMC '18), October 31-November 2, 2018, Boston, MA, USA}
\acmPrice{15.00}
\acmDOI{10.1145/3278532.3278574}
\acmISBN{978-1-4503-5619-0/18/10}

\begin{document}
\hypersetup{citecolor=,linkcolor=}
\title[Significance, Structure, and Stability of Internet Top Lists]{A Long Way to the Top: \\ Significance, Structure, and Stability of Internet Top Lists}
\author{Quirin Scheitle}
\affiliation{
  \institution{Technical University of Munich}
}

\author{Oliver Hohlfeld}
\affiliation{
  \institution{RWTH Aachen University}
}

\author{Julien Gamba}
\affiliation{
  \institution{IMDEA Networks Institute/\\Universidad Carlos III de Madrid}
}

\author{Jonas Jelten}
\affiliation{
  \institution{Technical University of Munich}
}

\author{Torsten Zimmermann}
\affiliation{
  \institution{RWTH Aachen University}
}

\author{Stephen D. Strowes}
\affiliation{
  \institution{RIPE NCC}
}

\author{Narseo Vallina-Rodriguez}
\affiliation{
  \institution{IMDEA Networks Institute/ICSI}
}

\renewcommand{\shortauthors}{Scheitle et al.}
\date{\today}


\begin{abstract}A broad range of research areas including Internet measurement, privacy, and network security rely on lists of target domains to be analysed; 
researchers make use of target lists for reasons of necessity or efficiency.
The popular Alexa list of one million domains is a widely used example. 
Despite their prevalence in research papers, 
the soundness of \toplists has
seldom been questioned by the community: 
little is known about the lists' creation, representativity, potential biases, stability, or overlap between lists.

In this study we survey the extent, nature, and evolution of \toplists used by
research communities.
We assess the structure and stability of these lists, and show that rank manipulation is possible for some lists. 
We also reproduce the results of several scientific studies to assess the impact of using
a top list at all, which list specifically, and the date of list creation. 
We find that 
\one top lists generally overestimate results compared to the general population by a significant margin, often even an order of magnitude, and
\two some top lists have surprising change characteristics, causing high day-to-day fluctuation and leading to result instability.
We conclude our paper with specific recommendations on the use of top lists, 
and how to interpret results based on top lists with caution.
\end{abstract}\vspace{-2mm}

 \maketitle
\section{Introduction}
\nocite{alexa, umbrella,majestic}
Scientific studies frequently make use of a sample of DNS domain names for various purposes, 
be it to conduct lexical analysis, to measure properties of domains, or to test whether a new algorithm works on real domains.
Internet \toplists, such as the Alexa or Cisco Umbrella \topm lists, serve the purpose of providing a reputedly representative sample of Internet domains in popular use.
These \toplists can be created with different methods and data sources, resulting in different sets of domains. 

The prevalence and opacity of these lists 
could have introduced an unchecked bias in science---for 10 networking venues in 2017 alone, we count 69 publications that use a top list.
This potential bias is based on the fact that curators of such top lists commonly conceal the data sources and ranking mechanism behind those lists, which are typically seen as a proprietary business asset in the search engine optimisation (SEO) space~\cite{majesticahrefs}.
This leaves researchers using those lists with little to no information about content, stability, biases, evolution and representativity of their contents.

In this work, we analyse three popular \toplists---Alexa Global~\cite{alexa}, Cisco Umbrella~\cite{umbrella}, and Majestic Million~\cite{majestic}---and discuss the following characteristics:

\para{Significance} In a survey of 687 net\-wor\-king-related papers published in 2017, we investigate if, to what extent, and for what purpose, these papers make use of Internet \toplists.
We find that 69 papers (10.0\%) make use of at least one \toplist (\cf~\Cref{sec:useoflists}).

\para{Structure} Domain properties in different \toplists, such as the surprising amount of invalid top-level domains (TLDs), low intersections between various lists (<30\%), and classifications of disjunct domains, are investigated in~\Cref{sec:structure}.

\para{Stability} We conduct in-depth longitudinal analyses of top list stability in \Cref{sec:stability}, revealing daily churn of up to 50\% of domains.

\para{Ranking Mechanisms} Through controlled experiments and reverse engineering of the Alexa toolbar, we shed light on the ranking mechanisms of different top lists. In one experiment, we place an unused test domain at a 22k rank in Umbrella (\cf~\Cref{sec:influencing}).

\para{Research Result Impact} Scientific studies that use \toplists for Internet research measure characteristics of the targets contained in each list, or the related infrastructure.
To show the bias inherent in any given target list, we run several experiments against \toplists and the general population of all \textit{com/net/org} domains.
We show that top lists significantly exaggerate results, and that results even depend on the day of week a list was obtained (\cf~\Cref{sec:impact}). 

We discuss related work and specific recommendations in~\Cref{sec:discussion}.

Throughout our work, we aim to adhere to the highest ethical standards, and aim for our work to be fully reproducible. 
We share code, data, and additional insights under
\begin{center}
  \url{https://toplists.github.io}
\end{center}

 \section{Domain Top Lists}\label{sec:background}

This section provides an overview of various domain \toplists and their
creation process. Each of these lists is updated daily. 

\para{Alexa}
The most popular and widely used \toplist is the Alexa Global \topm list\,\cite{alexa}.
It is generated based on web activity monitored by the Alexa browser plugin\footnote{Available for Firefox and Chrome. Internet Explorer discontinued June 2016\,\cite{alexatoolbarIE}}, ``directly measured sources''~\cite{alexapanel} and  ``over 25,000 different browser extensions''~\cite{alexamyths} over the past three months~\cite{alexalongtail} from ``millions of people''~\cite{alexapanel}.
No information exists on the plugin's user base, which opens questions on potential biases in terms of, \eg geography or age of its user base.
Alexa lists are generally offered for sale, with few free offerings.
Paid offerings include top lists per country, industry, or region.
The Global \topm list is the most popular free offering, available with no explicit license, and was briefly suspended in late 2016. 

\para{Cisco Umbrella}
Another popular \toplist is the Cisco Umbrella 1M, a \toplist launched in mid-December 2016.
This list contains the \topm domains (including subdomains) as seen by Cisco's OpenDNS service~\cite{umbrella}. 
This DNS-based nature is fundamentally different from collecting website visits or links.
Hence, the Umbrella list contains Fully Qualified Domain Names (FQDN) for any kind of Internet service, not just web sites as in the case of Alexa or Majestic.
Without explicit license, it is provided ``free of charge''.

\para{Majestic} 
The third \toplist is the Majestic Million~\cite{majestic}, released in October 2012.
This creative commons licensed list is based on Majestic's web crawler. 
It ranks sites by the number of \texttt{/24} IPv4-subnets linking to that site~\cite{majesticlaunch}. 
This is yet another data collection methodology and, similar to Alexa, heavily web-focused.
While the Majestic list is currently not widely used in research, we still include it in our study for its orthogonal mechanism, its explicitly open license, and its availability for several years.

\para{Other Top Lists} 
There are few other \toplists available, but as those are little used, not consistently available, or fluctuate in size, we will not investigate them in detail in this paper. 
Quantcast~\cite{quantcast} provides a list of the \topm most frequently visited websites per country, measured through their web intelligence plugin on sites. 
Only the US-based list can be downloaded; all other lists can only be viewed online and show ranks only when paid. 
The Statvoo~\cite{statvoo} list provides an API and a download for their \topm sites, but has frequently been inaccessible in the months before this publication. 
Statvoo does not offer insights about the metrics they use in their creation process. 
The Chrome UX report~\cite{chromeuxr} publishes telemetry data about domains popular with Chrome users. 
It does not, however, rank domains or provide a static-sized set of domains.
We also exclude the SimilarWeb Top Sites ranking~\cite{similarweb} as it is not available for free and little used in science.

 \section{Significance of Top Lists}\label{sec:useoflists}\label{sec:significance}

Scientific literature often harnesses one or more of the top lists outlined in
\Cref{sec:background}. 
To better understand how often and to what purpose top lists are used by
the literature, we survey 687 recent publications.

\subsection{Methodology}
\begin{table*}[t!]
\centering
\caption{\textbf{Left:} Use of top lists at 2017 venues. The `dependent' column
	indicates whether we deemed the results of the study to rely on the list used (`Y'), 
	or that the study relies on a list for verification (`V') of other results, or that a
	list is used but the outcome doesn't rely on the specific list selected (`N').
	The `date' column indicates how many papers stated the date of list download or measurement. \textbf{Right:} Type of lists used in 69 papers from left. Multiple counts for papers using multiple lists.}
\resizebox{1.65\columnwidth}{!}{
\begin{tabular}{l  c r rr rrrr c  c c }
	\toprule
	&  &  & \multicolumn{2}{c}{using list} & \multicolumn{3}{c}{\# dependent} & \multicolumn{2}{c}{\# date?} &  \\
	Venue & Area  & Papers & \# & \% $\downarrow$                       &   Y & V & N                & List & Study & References    \\
	\cmidrule(r){1-1}\cmidrule(r){2-2}\cmidrule(r{4pt}){3-3}\cmidrule(r){4-5}\cmidrule(r){6-8}\cmidrule(r){9-10}\cmidrule(r){11-11}
	ACM IMC		& Measurements	& 42 &	11 & 26.2\% &	 8 & 2 & 1 & 1 & 3 & \
			\cite{paper003:2017:imc, paper004:2017:imc, paper013:2017:imc, paper016:2017:imc,
				paper028:2017:imc, paper071:2017:imc, paper123:2017:imc, paper173:2017:imc,
				paper176:2017:imc, paper177:2017:imc, paper219:2017:imc}                  \\
	PAM		& Measurements	& 20 &	4 & 20.0\% &	 3 & 1 & 0 & 0 & 0 & \
			\cite{paper061:2017:pam, paper063:2017:pam, paper064:2017:pam, paper072:2017:pam} \\
	TMA		& Measurements	& 19 &	3 & 15.8\% &	 1 & 1 & 1 & 0 & 0 & \
			\cite{paper012:2017:tma, paper027:2017:tma, paper107:2017:tma}                    \\
\midrule
	USENIX Security	& Security	& 85 &	12 & 14.1\% &	 8 & 4 & 0 & 2 & 0 & \
			\cite{paper007:2017:usenixsec, paper014:2017:usenixsec, paper120:2017:usenixsec,
				paper122:2017:usenixsec, paper146:2017:usenixsec, paper170:2017:usenixsec,
				paper172:2017:usenixsec, paper179:2017:usenixsec, paper181:2017:usenixsec,
				paper182:2017:usenixsec, paper184:2017:usenixsec, paper232:2017:usenixsec}\\
	IEEE S\&P	& Security	& 60 &	5 & 8.3\% &	 3 & 2 & 0 & 1 & 1 & \
			\cite{paper011:2017:ieeesp, paper018:2017:ieeesp, paper106:2017:ieeesp,
				paper144:2017:ieeesp, paper208:2017:ieeesp, paper229:2017:ieeesp}         \\
	ACM CCS		& Security	& 151 &	11 & 7.3\% &	 4 & 5 & 2 & 1 & 1 & \
			\cite{paper104:2017:ccs, paper169:2017:ccs, paper174:2017:ccs, paper175:2017:ccs,
				paper183:2017:ccs, paper185:2017:ccs, paper207:2017:ccs, paper216:2017:ccs,
				paper220:2017:ccs, paper230:2017:ccs, paper231:2017:ccs}                  \\
	NDSS		& Security	& 68 &	3 & 4.4\% &	 2 & 0 & 1 & 0 & 0 & \
			\cite{paper121:2017:ndss, paper206:2017:ndss, paper217:2017:ndss}                 \\
\midrule
	ACM CoNEXT	& Systems	& 40 &	4 & 10.0\% &	 2 & 1 & 1 & 0 & 1 & \
			\cite{paper005:2017:conext, paper008:2017:conext, paper015:2017:conext,
				paper029:2017:conext, paper145:2017:conext} \\
	ACM SIGCOMM	& Systems	& 38 &	3 & 7.9\% &	 3 & 0 & 0 & 0 & 0 & \
			\cite{paper006:2017:sigcomm, paper017:2017:sigcomm, paper062:2017:sigcomm}        \\
\midrule
	WWW		& Web Tech.	& 164 &	13 & 7.9\% &	11 & 1 & 1 & 2 & 3 & \
			\cite{paper030:2017:www, paper060:2017:www, paper105:2017:www, paper124:2017:www,
				paper134:2017:www, paper143:2017:www, paper171:2017:www, paper204:2017:www,
				paper205:2017:www, paper209:2017:www, paper214:2017:www, paper215:2017:www,
				paper218:2017:www}                                                        \\		
	\midrule
	Total		&		& 687 &	69  & 10.0\% &	45 & 17 & 7 & 7 & 9 &         \\
	\bottomrule
\end{tabular}}
\hfill
\resizebox{.405\columnwidth}{!}{
	\begin{tabular}{lr|lr}
	\toprule
	\multicolumn{4}{c}{Alexa Global Top \ldots} \\
	\midrule
	1M & 29 & 5k & 2 \\
	100k & 2 &  1k & 5 \\
	75k & 1  & 500 & 8  \\
	50k & 2  & 400 & 1 \\
	25k & 2 & 300 & 1  \\
	20k & 1 & 200 & 1  \\
	16k & 1 & 100 & 8  \\
	10k & 11 & 50 & 3\\
	8k & 1 & 10 & 1 \\
	\midrule
	\multicolumn{3}{l}{Alexa Country:} & 2 \\	
	\multicolumn{3}{l}{Alexa Category: } & 2 \\	
	\multicolumn{3}{l}{Umbrella 1M:} & 3 \\		
	\multicolumn{3}{l}{Umbrella 1k: } & 1  \\		
	\bottomrule
\end{tabular}
}\label{tab:conferencecoverage}
\end{table*}

We survey papers published at 10 network-related venues in 2017, listed in \Cref{tab:conferencecoverage}.  
First, we search the 687 papers published at these venues for keywords\footnote{Those being: ``alexa'', ``umbrella'', and ``majestic''} in an automated manner.
Next, we inspect matching papers manually to remove false positives (\eg Amazon's Alexa home assistant, or an author named Alexander), or  papers that mention the lists without actually using them as part of a study.

Finally, we reviewed the remaining 69 papers (10.0\%) that made use of a top list, with various aims in mind: 
to understand the top lists used (\Cref{subsec:study-ranks-used}), the nature of the study and the
technologies measured (\Cref{subsec:study-types}), whether the study was
dependent on the list for its results (\Cref{subsec:study-dependent}),
and whether the study was possibly replicable
(\Cref{subsec:study-replicable}). 
Table~\ref{tab:conferencecoverage} provides an overview of the results.

We find the field of Internet measurement to be most reliant on \toplists, used in 22.2\% of the surveyed papers.
Other fields also use \toplists frequently, such as security (8.5\%), systems
(6.4\%) and web technology (7.9\%).

\subsection{Top Lists Used}\label{subsec:study-ranks-used}

We first investigate which lists and what subsets of lists are typically used; 
Table~\ref{tab:conferencecoverage} provides an overview of lists used in the studies we identified.
We find 29 studies using the full Alexa Global \topm, 
the most common choice among inspected publications, followed by a surprising variety of Alexa \topm subsets (\eg Top\,10k). 

All papers except one~\cite{paper006:2017:sigcomm} use a list collated by Alexa. 
This paper instead uses the Umbrella \toph list to assess importance of ASes showing BGP bursts. 
No paper in our study used the Majestic list.

A study may also use multiple distinct subsets of a list.
For example, one study uses the Alexa Global \topk,
10K, 500K and \topm at different stages of the study~\cite{paper121:2017:ndss}.
We count these as distinct use-cases in the right section of Table~\ref{tab:conferencecoverage}.

We also find that 59 studies exclusively use Alexa as a source for domain names.
Ten papers use lists from more than one origin; one paper uses the Alexa Global \topm, the Umbrella \topm, and various DNS zone files as sources~\cite{paper173:2017:imc}.
In total, two studies make use of the Cisco Umbrella \topm~\cite{paper029:2017:conext, paper173:2017:imc}.

Category and country-specific lists are also being used: 
eight studies use country-specific lists from Alexa, usually choosing only one country; one
study selected 138 countries~\cite{paper063:2017:pam}. 
Category-based lists are rarer still: two studies made use of category subsets~\cite{paper016:2017:imc, paper062:2017:sigcomm}.

\subsection{Characterisation of Studies}\label{subsec:study-types}
To show that \toplists are used for various types of studies,
we look at the range of topics covered and technologies measured in our surveyed papers.
For each paper we assigned a broad \textit{purpose}, and the network \textit{layer} in focus. 

\para{Purposes} 
For all papers, we reviewed the broad area of study.
The largest category we
identified encompasses various aspects of security, across 38 papers in total: this includes phishing
attacks~\cite{paper209:2017:www, paper214:2017:www}, session safety during
redirections~\cite{paper215:2017:www}, and
domain squatting~\cite{paper220:2017:ccs}, to name a few.
Nine more papers study aspects of privacy \& censorship,
such as the Tor overlay network~\cite{paper121:2017:ndss}, or user
tracking~\cite{paper122:2017:usenixsec}.
Network or application performance is also a popular area: ten papers in our survey
focus on this, \eg~HTTP/2 server push~\cite{paper030:2017:www}, mobile web
performance~\cite{paper062:2017:sigcomm}, and Internet
latency~\cite{paper063:2017:pam}.
Other studies look at economic aspects such as hosting providers.

\para{Layers} We also reviewed the network layers measured in each study.
Many of the papers we surveyed focus on web infrastructure:
22 of the papers are concerned with content, 8 focus on the HTTP(S)
protocols, and 7 focus on applications (\eg browsers~\cite{paper179:2017:usenixsec, paper181:2017:usenixsec}).

Studies relating to core network protocols are commonplace:
DNS~\cite{paper007:2017:usenixsec, paper121:2017:ndss, paper146:2017:usenixsec, paper169:2017:ccs, paper174:2017:ccs}
(we identified 3 studies relating to domain names as separate from
DNS \emph{protocol} measurements~\cite{paper217:2017:ndss, paper219:2017:imc, paper220:2017:ccs}), 
TCP~\cite{paper071:2017:imc, paper107:2017:tma}, and 
IP~\cite{paper003:2017:imc, paper004:2017:imc, paper005:2017:conext, paper006:2017:sigcomm, paper028:2017:imc, paper027:2017:tma},
and TLS/HTTPS~\cite{paper104:2017:ccs, paper134:2017:www, paper170:2017:usenixsec, paper172:2017:usenixsec, paper173:2017:imc, paper215:2017:www, paper216:2017:ccs} layer measurements are common in our survey.

Finally, we identify 12 studies whose experimental design measures more than one specific
layer; \eg cases studying a full connection establishment (from initial DNS query to HTTP request). 

We conclude from this that top lists are frequently used to explicitly or implicitly measure DNS, IP, and TLS/HTTPS characteristics, which we investigate in depth in \Cref{sec:impact}.

\subsection{Are Results Dependent on Top Lists?}\label{subsec:study-dependent}

In this section, we discuss how dependent study results are on top lists.
For this, we fill the ``dependent'' columns in \Cref{tab:conferencecoverage} as follows:

\para{Dependent (Y)} 
Across all papers surveyed, we identify 45 studies whose results may be affected by the list chosen. 
Such a study would take a list of a certain day, measure some characteristic over the set of domains in that list, and draw conclusions about the measured characteristic. 
In these cases, we say that the results \textit{depend} on the list being used: 
a different set of domains in the list may have yielded different results.

\para{Verification (V)}
We identify 17 studies that use a list only to verify their results.
A typical example may be to develop some algorithm to find domains with a certain property, and then use a top list to check whether these domains are popular.
In such cases, the algorithm developed is independent of the list's content.

\para{Independent (N)}
Eight studies cite and use a list, but we determine that their results are not necessarily reliant on the list. 
These papers typically use a \toplist as one source among many, such that changes in the \toplist would likely not affect the overall results.

\subsection{Are Studies Replicable?}\label{subsec:study-replicable}

Repeatability, replicability, and reproducibility are ongoing concerns in Computer Networks~\cite{AcmArtifacts, reproduc2017} and Internet Measurement~\cite{Flittner}.
While specifying the date of when a top list was downloaded, and the date when measurements where conducted, are not necessarily sufficient to reproduce studies, they are important first steps.

Table~\ref{tab:conferencecoverage} lists two ``date'' columns that indicate whether the list download date or the measurement dates were given\footnote{We require a specific day to be given to count a paper, the few papers just citing a year or month were counted as no date given}. 
Across all 69 papers using top lists, only 7 stated the date the list was retrieved, and 9 stated the measurement date.
Unfortunately, only 2 papers give both the list and measurement data and hence fulfil these basic criteria for reproducibility.
This does not necessarily mean that the other papers are not reproducible, authors may publish the specific top list used as part of data, or authors might be able to provide the dates or specific list copies upon inquiry. 
However, recent investigations of reproducibility in networking hints that this may be an unlikely expectation~\cite{Flittner, saucez2018thoughts}.
We find two papers that explicitly discuss instability and bias of top lists, and use aggregation or enrichment to stabilise results~\cite{paper018:2017:ieeesp, paper029:2017:conext}.

\subsection{Summary}
Though our survey has a certain level of subjectivity, we consider its broad findings meaningful:
\one that top lists are frequently used,
\two that many papers' results depend on list content, and
\three that few papers indicate precise list download or measurement dates. 

We also find that the use of \toplists to measure network and security characteristics (DNS, IP, HTTPS/TLS) is common. 
We further investigate how \toplist use impacts result
quality and stability in studies by measuring these layers in \Cref{sec:impact}.

 \section{Top Lists Dataset}
For the three lists we focus on in this study, we source daily snapshots as far back as possible. Many
snapshots come from our own archives, and others were shared with us by other members of the research community, such as \cite{allman2018robustness,wahlisch2015ripki,holzimc2011}.
\Cref{tab:data} gives an overview of our datasets along with some metrics discussed in \Cref{sec:structure}.
For the Alexa list, we have a dataset with daily snapshots from January 2009 to March 2012, named \textit{AL0912}, and another from April 2013 to April 2018, named \textit{AL1318}. The Alexa list underwent a significant change in January 2018;
for this we created a partial dataset named \textit{AL18} after this change.
For the Umbrella list, we have a dataset spanning 2016 to 2018, named \textit{UM1618}.
For the Majestic Million list, we cover June 2017 to April 2018. 

As many of our analyses are comparative between lists, we create a JOINT dataset, spanning the overlapping period from June 2017 to the end of April 2018. 
We also sourced individual daily snapshots from the community and the Internet Archive~\cite{archivealexacrawl}, but only used periods with continuous daily data for our study.

 \section{Structure of Top Lists}\label{sec:structure}
In this section, we analyse the structure and nature of the three top lists in our study.
This includes questions such as top level domain (TLD) coverage, subdomain depth, and list intersection.

\textbf{DNS Terms} used in this paper, for clarity, are the following:
for \textit{www.net.in.tum.de}, \textit{.de} is the public suffix\footnote{per Public Suffix List~\cite{pslgithub}, a browser-maintained list aware of cases such as \textit{co.uk.}} (and top level domain), 
\textit{tum.de} is the base domain, 
\textit{in.tum.de} is the first subdomain, and \textit{net.in.tum.de} is the second subdomain.
Hence, \textit{www.net.in.tum.de} counts as a third-level subdomain.
\begin{table*}[t]
	\centering
	\caption{Datasets: mean of valid TLDs covered ($\mu_{TLD}$), mean of base domains ($\mu_{BD}$), mean of sub-domain level spread ($SD_n$ for share of n-th level subdomains, $SD_M$ for maximum sub-domain level), mean of domain aliases ($DUP_{SLD}$), mean of daily change ($\mu_\Delta$) and mean of new (\ie not included before) domains per day ($\mu_{NEW}$).  {\small Footnote 4: Average after Alexa's change in January 18.}}	
	\vspace{-3mm}
	\resizebox{\textwidth}{!}
	{
		\begin{tabular}{lllrll r lr  r lrr}
			\toprule
			List                             & Top &  Dataset &    Dates & $\mu_{TLD}\pm\sigma$ & $\mu_{BD}\pm\sigma$ & $SD_1$ & $SD_2$ & $SD_3$ & $SD_{M}$ & $DUP_{SLD}$ &  $\mu_\Delta$ &  $\mu_{NEW}$ \\ \midrule
			Alexa & 1M  & AL0912 &   29.1.09--16.3.12 & 248 $\pm$ 2         &       973k $\pm$ 2k &    1.6\%   &    0.4\%   &$\approx$0\% &       4  &  47k $\pm$ 2k &  23k & n/a \\
			Alexa& 1M  & AL1318 &   30.4.13--28.1.18 & 545 $\pm$ 180         &       972k $\pm$ 6k &    2.2\%   &    0.1\%   &$\approx$0\% &       4  &  49k $\pm$ 3k & 21k &  5k \\
			Alexa &1M  & AL18 &   29.1.18--30.4.18 & 771 $\pm$ 8         &       962k $\pm$ 4k &    3.7\%   &  $\approx$0\%  &$\approx$0\% &       4  &  45k $\pm$ 1k & 483k  & 121k \\			
			\midrule
			Alexa& 1M   & JOINT &   6.6.17--30.4.18 & 760 $\pm$ 11         &       972k $\pm$ 7k &    2.6\%   &    $\approx$0\%   &$\approx$0\% &       4  &  51k $\pm$ 4k & 147k  & 38k \\

			Umbrella& 1M & JOINT & 6.6.17--30.4.18 & 580 $\pm$ 13       &      273k $\pm$ 13k &  49.9\%     &  14.7\%     & 5.9\% &    33     &  15k $\pm$ 1k&  100k &  22k \\
			Majestic& 1M & JOINT &   6.6.17--30.4.18 & 698 $\pm$ 14         &      994k $\pm$ 617 &   0.4\%   &  $\approx$0\%     &$\approx$0\% &   4      & 49k $\pm$ 1k &  6k & 2k  \\ 
			\midrule
			Alexa& 1k  & JOINT &   6.6.17- 30.4.18 & 105 $\pm$ 3  		& ~~990  $\pm$2 & 1.3\% & 0.0\% & 0.0\%  & 1 				& ~~~22 $\pm$ 2  & 9 (78\protect\footnotemark[4]) & 4 (8\protect\footnotemark[4]) \\						
			Umbrella &1k  & JOINT & 6.6.17--30.4.18 & ~~13 $\pm$ 1  & ~~317 $\pm$6 & 52.0\% & 14\% & $\approx$0\%  & 6 & ~~~11 $\pm$ 2  & 44 & 2 \\			
			Majestic &1k & JOINT &   6.6.17--30.4.18 & ~~50 $\pm$ 1  &  ~~939 $\pm$3 & 5.9\% & 0.1\% & 0.1\% &4 					& ~~~32 $\pm$ 1 & 5  & .8 \\ 			
			\midrule
			Umbrella &1M & UM1618 & 15.12.16--30.4.18 & 591 $\pm$ 45       &      281k $\pm$ 16k &  49.4\%     &  14.5\%     & 5.7\% &    33     &  15k$\pm$1k&  118k & n/a \\
			\bottomrule& & & 
		\end{tabular}
	}\vspace{-3mm}
	\label{tab:data}
\end{table*}

 \subsection{Domain Name Depth and Breadth}\label{subsec:structure:tld}
A first characteristic to understand about top lists is the scope of their coverage: 
how many of the active TLDs do they cover, and how many do they miss?
How deep are they going into specific subdomains, choosing trade-offs between breadth and depth?

\textbf{TLD Coverage} is a first indicator of list breadth. Per IANA~\cite{ianatld,terminatedtlds}, 1,543 TLDs exist as of May 20th, 2018.
Based on this list, we count valid and invalid TLDs per list. 
The average coverage of valid TLDs in the JOINT period is $\approx$700 TLDs, covering only about 50\% of active TLDs.
This implies that measurements based on top lists may miss up to 50\% of TLDs in the Internet.

At the \topk level we find quite different behaviour with 105 valid TLDs for Alexa, 50 for Majestic, but only 13 (\textit{com/net/org} and few other TLDs) for Umbrella. 
We speculate that this is rooted in DNS administrators from highly queried DNS names preferring the smaller set of professionally managed and well-established top level domains over the sometimes problematic new gTLDs~\cite{stopusingio,ioerror, ioregistrar}.

Invalid TLDs occur neither in any \topk domains nor in the Alexa \topm domains, but as a minor count in the Majestic \topm (7 invalid TLDs, resulting in 35 domain names), and significant count in the Umbrella \topm: there, we can find 1,347 invalid TLDs\footnote{Examples for invalid TLDs:  \textit{instagram}, \textit{localdomain}, \textit{server}, \textit{cpe}, \textit{0}, \textit{big}, \textit{cs}}, in a total of 23k domain names (2.3\% of the list).
This is an early indicator of a specific characteristic in the Umbrella list: 
invalid domain names queried by misconfigured hosts or outdated software can easily get included into the list.

Comparing valid and invalid TLDs also reveals another structural change in the Alexa list on July 20th, 2014:
before that date, Alexa had a fairly static count of 206 invalid and 248 valid TLDs.
Perhaps driven by the introduction of new gTLDs from 2013~\cite{newtldtimelines}, Alexa changed its filtering:
After that date, invalid TLDs have been reduced to $\approx$0, and valid TLDs have shown continued growth from 248 to $\approx$800.
This confirms again that top lists can undergo rapid and unannounced changes in their characteristics, which may significantly influence measurement results.

\textbf{Subdomain Depth} is an important property of top lists. Base domains offer more breadth and variety in setups, while subdomains may offer interesting targets besides a domain's main web presence. 
The ratio of base to subdomains is hence a breadth/depth trade-off, which we explore for the three lists used.
\Cref{tab:data} shows the average number of base domains ($\mu_{BD}$) per top list. 
We note that Alexa and Majestic contain almost exclusively base domains with few exceptions (\eg for \texttt{blogspot}).
In contrast, 28\% of the names in the Umbrella list are base domains, \ie Umbrella emphasises depth of domains.
\Cref{tab:data} also details the subdomain depth for a single-day snapshot (April 30, 2018) of all lists.
As the Umbrella list is based on DNS lookups, such deep DNS labels can easily become part of the Umbrella list, regardless of the origin of the request. 
In fact, Umbrella holds subdomains up to level 33 (\eg domains with extensive \textit{www} prefixes or `.'-separated OIDs).

We also note that the base domain is usually part of the list when its subdomains are listed. 
On average, each list contains only few hundred subdomains whose base domain is not part of the list.

\textbf{Domain Aliases} are domains with the same second-level domain, but different top-level domains, \eg \textit{google.com} and \textit{google.de}.
\Cref{tab:data} shows the number of domain aliases as $DUP_{SLD}$. 
We find a moderate level of $\approx$5\% of domain aliases within various top lists, with only 1.5\% for Majestic. 
Analysis reveals a very flat distribution, with the top entry \textit{google} at $\approx$200 occurrences.

\subsection{Intersection between Lists}
\begin{figure*}[h]
	\href{https://toplists.github.io/#intersect}{	\begin{subfigure}[t]{0.32\textwidth}
		\includegraphics[width=\columnwidth]{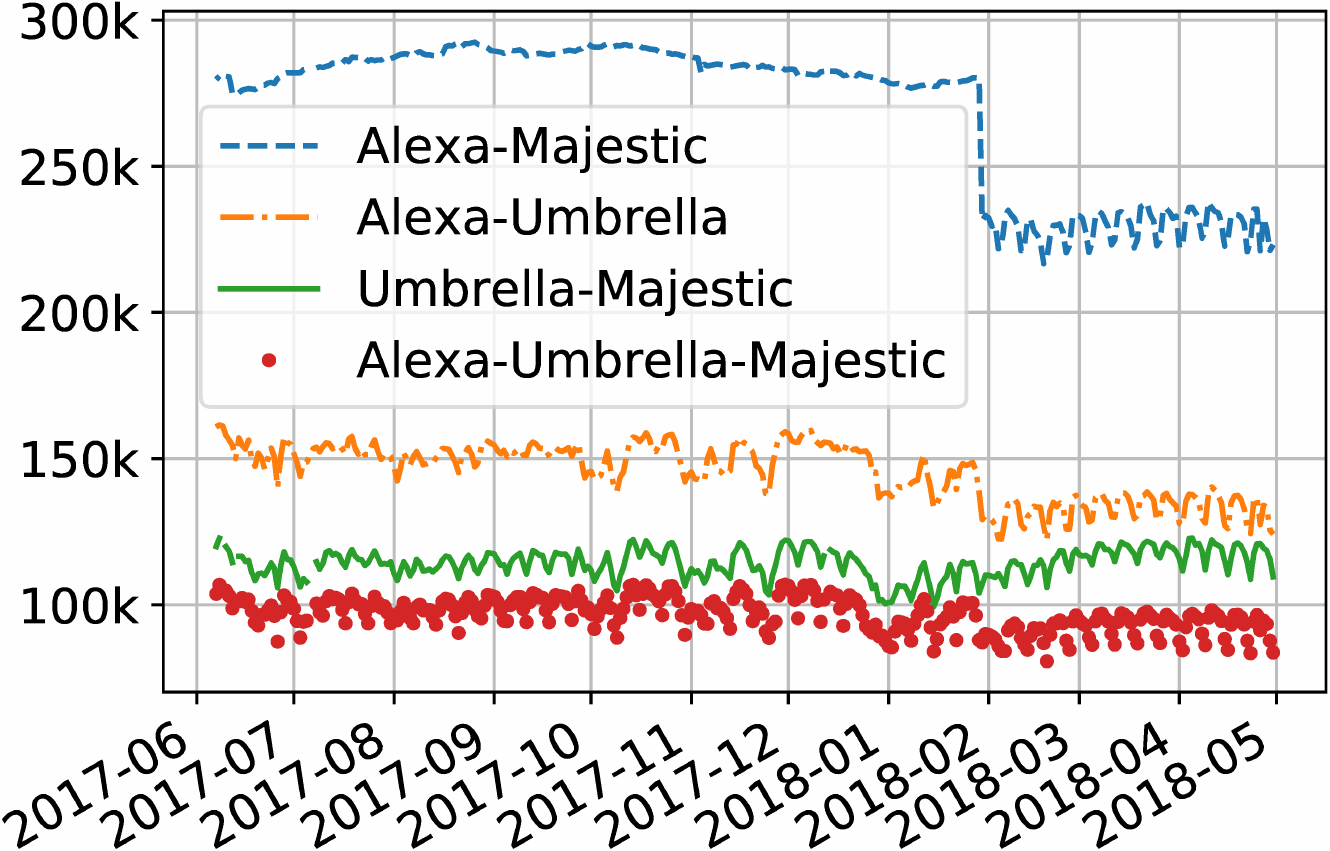} 		\vspace{-5mm}
		\caption{Intersection between Top1M lists \textit{(live)}.}
		\label{subfig:intersection}
	\end{subfigure}}
	\hfill
	\href{https://toplists.github.io/#dailychange}{	\begin{subfigure}[t]{0.32\textwidth}
		\includegraphics[width=\columnwidth]{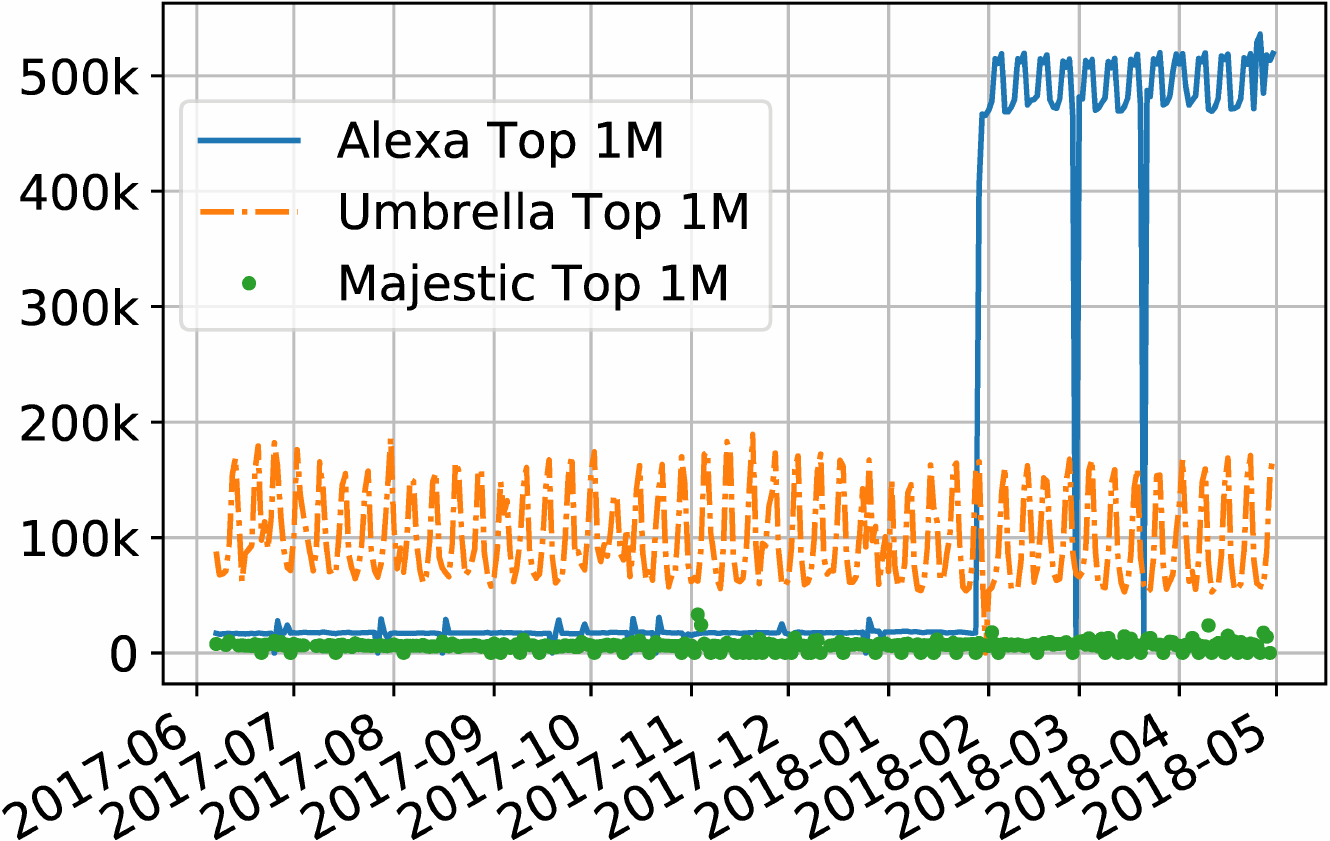} 		\vspace{-5mm}
		\caption{Daily changes of Top1M entries.\textit{(live)}}
		\label{subfig:dailychange}
	\end{subfigure}}
	\hfill
	\begin{subfigure}[t]{0.33\textwidth}
		\includegraphics[width=\columnwidth]{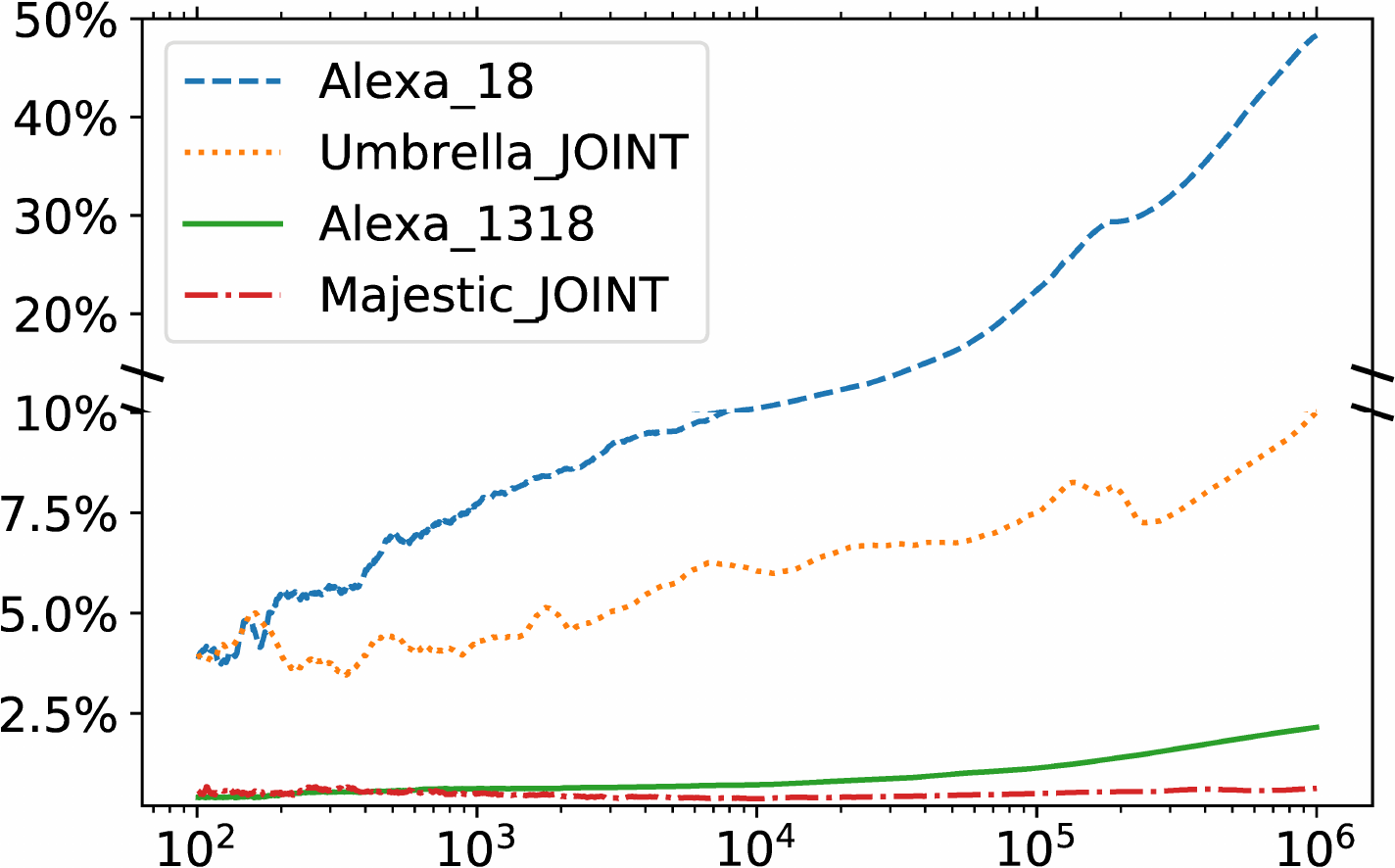} 		\vspace{-5mm}       
		\caption{Average \% daily change over rank.}		\label{subfig:dailychangerank}
	\end{subfigure}
	\vspace{-3mm}
	\caption{Intersection, daily changes and average stability of top lists (y-axis re-scaled at 10\% in right plot). \textit{Click for live version/source code}}
	\vspace{-3mm}   
\end{figure*}

We next study intersection between lists---all 3 lists in our study promise a view on the most popular  domains (or websites) in the Internet, hence measuring how much these lists agree\footnote{To control for varying subdomain length, we first normalise all lists to unique base domains (cf. $\mu_{BD}$ in \Cref{tab:data}, reducing \eg Umbrella to 273k base domains)} is a strong indicator of bias in list creation.
~\Cref{subfig:intersection} shows the intersection between top lists over time during the JOINT period. We see that the intersection is quite small:
for the Top1M domains, Alexa and Majestic share 285k domains on average during the JOINT duration.
Alexa and Umbrella agree on 150k, Umbrella and Majestic on 113k, and all three only on 99k out of 1M domains.

For the Top1k lists, the picture is more pronounced. 
On average during the JOINT period, Alexa and Majestic agree on 295 domains, 
Alexa and Umbrella on 56, Majestic and Umbrella on 65, and all three only on 47 domains.

This disparity between top domains suggests a high bias in the list creation.
We note that even both web-based lists, Alexa and Majestic, only share an average of 29\% of domains.

Standing out from Figure~\ref{subfig:intersection} is the fact that the Alexa list has changed 
its nature in January 2018, reducing the average intersection with Majestic from 285k to 240k.
This change also introduced a weekly pattern, which we discuss further in ~\Cref{sec:weekly}. 
We speculate that Alexa might have reduced its 3-month sliding window~\cite{alexalongtail}, making the list more volatile and susceptible to weekly patterns.
We contacted Alexa about this change, but received no response.

\subsection{Studying Top List Discrepancies}\label{subsec:discrepancies}
The low intersection between Umbrella and the other lists could be rooted in the DNS vs. web-based creation. 
Our hypothesis is that the web-based creation of Alexa and Majestic lists tends to miss domains providing embedded content as well as domains popular on mobile applications~\cite{razaghpanah2018apps,paper029:2017:conext}.
In this section, we explore the origin of discrepancies across domain lists.

We aggregate the Alexa, Umbrella, and Majestic \topk domains from the 
last week of April 2018, and analyse the set of 3,005 disjunct domains across 
these lists, \ie those found only in a single list.
40.7\% of these domains originate from Alexa, 37.1\% from Umbrella, and 22.1\% from Majestic. Subsequently, we identify whether the disjunct domains are associated with mobile traffic or third-party advertising and tracking services not actively visited by users, but included through their DNS lookups.
We opt against utilizing domain classifiers such as 
the OpenDNS Domain Tagging service~\cite{opendnsdomaintagging}, as it has been reported
that categories are vague and coverage is low~\cite{razaghpanah2018apps}.

Instead, we use the data captured by the Lumen Privacy 
Monitor~\cite{razaghpanah2015haystack} to associate 
domains with mobile traffic for more than 60,000 Android apps, 
and use popular anti-tracking blacklists such as 
MalwareBytes' hpHosts ATS file~\cite{hphosts}. We also check if the domains from a given \toplist can be found in the aggregated \topm of the
other two \toplists during the same period of time.
Table~\ref{tab:domains_classification} summarises the results.
As we suspected, Umbrella has significantly more domains flagged as ``mobile traffic'' and third-party advertising and tracking services than the other lists. 
It also has the lowest proportion of domains shared with other \topm lists.

This confirms that Umbrella is capable of capturing domains from any device using OpenDNS, such as mobile and IoT devices, and also include domains users are not aware of visiting, such as embedded third-party trackers in websites. 
Alexa and Majestic provide a web-specific picture of popular Internet domains.

\begin{table}
  \small
  \centering
  \caption{Share of one-week \topk disjunct domains present in
	hpHosts (blacklist), Lumen (mobile), and \topm of other \toplists.}
	\vspace{-3mm}
  \begin{tabular}{l r r r r}
    \toprule
    List & \# Disjunct  & \%\,hpHosts & \%\,Lumen & \%\,\topm \\
    \midrule
    Alexa & 1,224 & 3.10\% & 1.55\% & 99.10\% \\
    Umbrella & 1,116  & 20.16\% & 39.43\% & 25.63\% \\
    Majestic & 665 & 1.95\% & 3.76\% & 93.63\% \\
    \bottomrule
  \end{tabular}
  \label{tab:domains_classification}
  \vspace{-3mm}
\end{table}

 \section{Stability of Top Lists}\label{sec:stability}

Armed with a good understanding of the \textit{structure} of top lists, we now focus on their stability over time. 
Research has revealed hourly, daily and weekly patterns on ISP traffic and service load, as well as significant regional and demographic differences in accessed content due to user habits~\cite{lakhina2004structural,papagiannaki2003long,gill2007youtube,cha2007tube}. 
We assess whether such patterns also manifest in \toplists, as a first step towards understanding the impact of studies selecting a \toplist at a given time.

\subsection{Daily Changes}

We start our analysis by understanding the composition and evolution of top lists on a daily basis. 
As all top lists have the same size, we use the raw count of daily changing domains for comparison. 

~\Cref{subfig:dailychange} shows the count of domains that were removed daily, specifically the count of domains present in a list on day $n$ but not on day $n$+$1 $.
The Majestic list is very stable (6k daily change), the Umbrella list offers 
significant churn (118k), and the Alexa list used to be stable (21k), 
but drastically changed its characteristic 
in January 2018 (483k), becoming the most unstable list. 

The fluctuations in the Umbrella list, and in the Alexa list after January 2018, are weekly patterns,
which we investigate closer in \Cref{sec:weekly}.
The average daily changes are given in column $\mu_\Delta$ of \Cref{tab:data}.

\textbf{Which Ranks Change?} 
Previous studies of Internet traffic revealed that the distribution of accessed domains and services follows a power-law distribution~\cite{paper145:2017:conext,lakhina2004structural,papagiannaki2003long,gill2007youtube,cha2007tube}.
Therefore, the ranking of domains in the long tail should be based on significantly smaller and hence less reliable numbers. 

~\Cref{subfig:dailychangerank} displays the stability of lists depending on subset size.
The y-axis shows the mean number of daily changing domains in the top\,X domains, where X is depicted on the x-axis. 
For example, an x-value of 1000 means that the lines at this point show the average daily change per list for the \topk domains.
The figure shows instability increasing with higher ranks for Alexa and Umbrella, but not for Majestic.
We plot Alexa before and after its January 2018 change, highlighting the significance of the change across all its ranks--even its \topk domains have increased their instability from 0.62\% to 7.7\% of daily change. 

\begin{figure*}
	\begin{subfigure}[t]{0.3\textwidth}
		\includegraphics[width=\columnwidth, height=0.7\columnwidth, keepaspectratio]{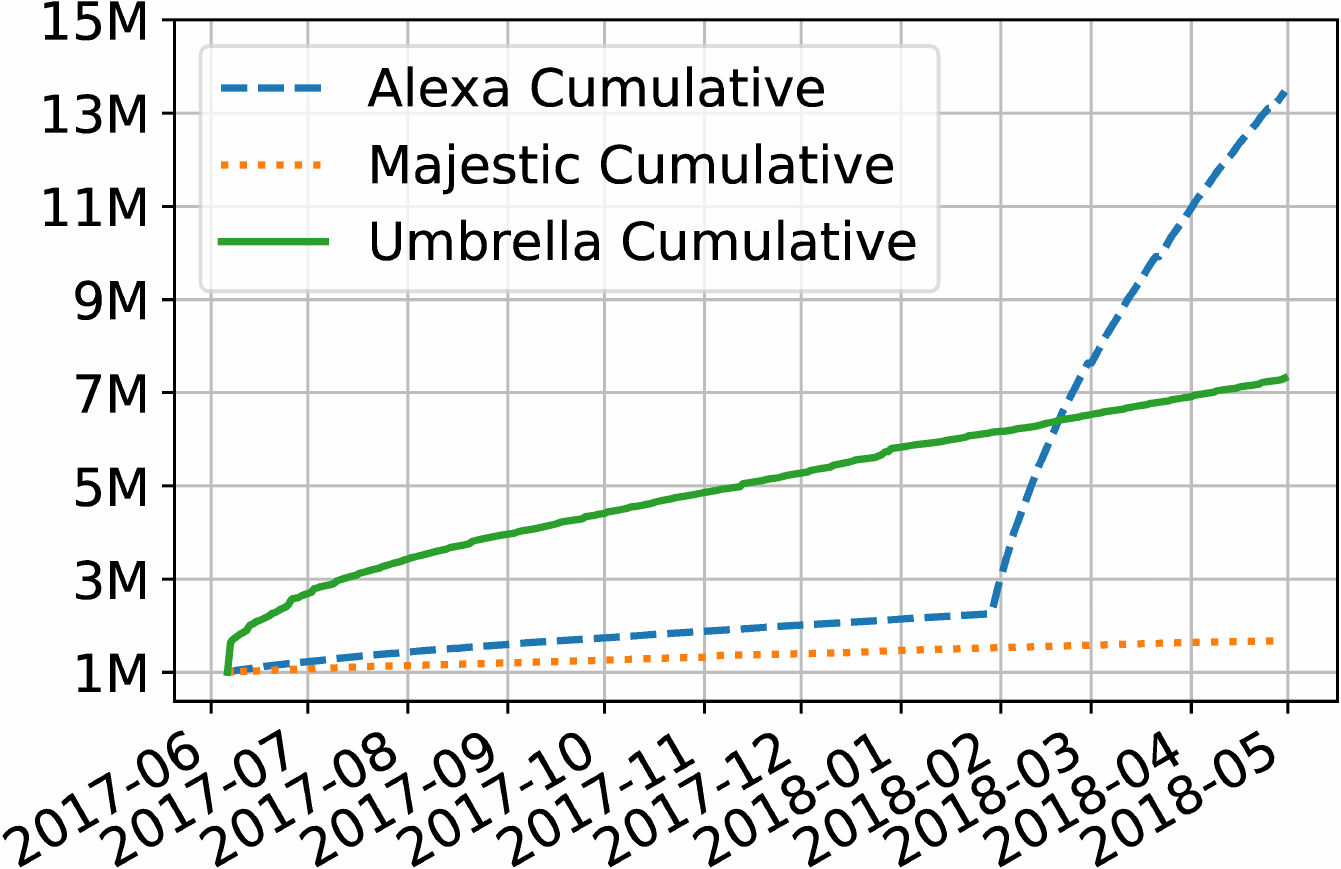} 		\vspace{-5mm}
		\caption{Cumulative sum of all domains ever included in \topm lists (\topk similar).}
		\label{subfig:runup}
	\end{subfigure}
	\hfill
	\begin{subfigure}[t]{0.3\textwidth}
		\includegraphics[width=\columnwidth]{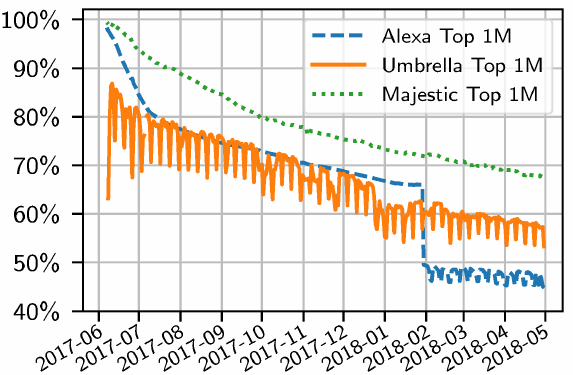}
		\vspace{-5mm}
    \caption{List intersection against a fixed starting set (median value
    of seven different starting days)}
		\label{subfig:moving_days}
	\end{subfigure}
	\hfill
	\begin{subfigure}[t]{0.3\textwidth}
		\includegraphics[width=\columnwidth, height=0.7\columnwidth, keepaspectratio]{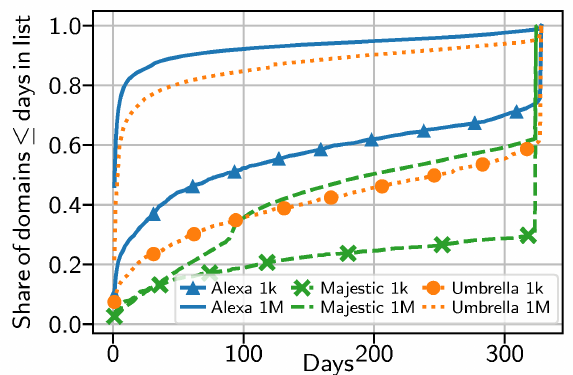}
		\vspace{-5mm}
		\caption{CDF of \% of domains over days included in \topm and \topk lists.}
		\label{subfig:ndays}
	\end{subfigure}
	\hfill
	\vspace{-3mm}
	\caption{Run-up and run-down of domains; share of days that a domains spend in a top list for the JOINT dataset.}
	\vspace{-3mm}	
\end{figure*}

\textbf{New or In-and-out Domains?} Daily changes in \toplists 
may stem from new domains joining, or from previously contained domains re-joining. To evaluate this, we cumulatively sum all the unique domains ever seen in a list in~\Cref{subfig:runup}, \ie a list with only permutations of the same set of domains would be a flat line.
Majestic exhibits linear growth: every day, about 2k previously not included domains are added to it --- approximately a third of the 6k total changing domains per day (\ie 4k domains have rejoined). Over the course of a year, the total count of domains included in the Majestic list is 1.7M. 
Umbrella adds about 20k new domains per day (out of 118k daily changing domains), resulting in 7.3M domains after one year. Alexa grows by 5k (of 21k) and 121k (of 483k) domains per day, before and after its structural change in January 2018. 
Mainly driven from the strong growth after Alexa's change, its cumulative number of domains after one year is 13.5M. 
This means that a long-term study of the Alexa \topm will, over the course of this year, have measured 13.5M distinct domains.

Across all lists, we find an average of 20\% to 33\% of daily changing domains to be new domains, \ie entering the list for the first time. 
This also implies that 66\% to 80\% of daily changing domains are domains that are repeatedly removed from and inserted into a list. 
We also show these and the equivalent \topk numbers in column $\mu_{NEW}$ of \Cref{tab:data}.

This behaviour is further confirmed in~\Cref{subfig:moving_days}. In this
figure, we compute the intersection between a fixed starting day and the
upcoming days. We compute it seven times, with each day of the first week of
the JOINT dataset as the starting day. \Cref{subfig:moving_days} shows
the daily median value between these seven intersections.

This shows several interesting aspects:
\one the long-term trend in temporal decay per list, confirming much of what we have seen before (high stability for Majestic, weekly patterns and high instability for Umbrella and the late Alexa list)
\two the fact that for Alexa and Umbrella, the decay is non-monotonic, \ie a set of domains is leaving and rejoining at weekly intervals.

\begin{figure*}[t]
	\captionsetup[subfigure]{width=0.975\textwidth, justification=centering}
	\begin{subfigure}[t]{.33\textwidth}
		\includegraphics[width=\columnwidth]{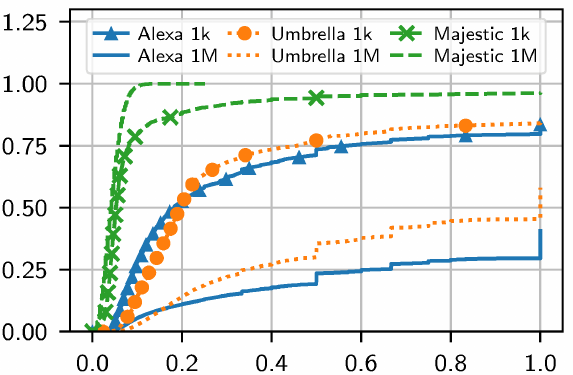}
		\vspace{-5mm}
		\caption{\small Kolmogorov-Smirnov (KS) distance between weekend and weekday distributions.}		\label{subfig:cdf_distance}
	\end{subfigure}	\begin{subfigure}[t]{.33\textwidth}
		\includegraphics[width=\columnwidth]{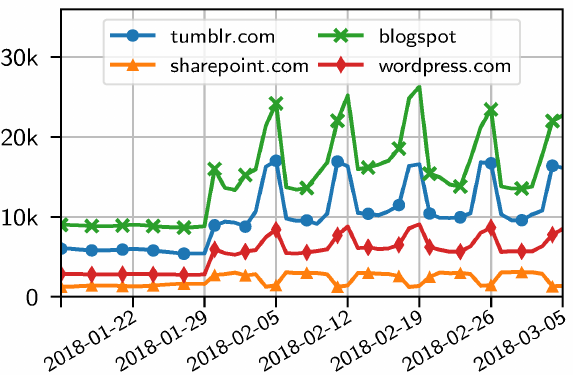}
		\vspace{-5mm}       
    \caption{Weekday/weekend dynamics in Alexa \topm Second-Level-Domains (SLDs).}		\label{subfig:slds_alexa}
	\end{subfigure}	\begin{subfigure}[t]{.33\textwidth}
		\includegraphics[width=\columnwidth]{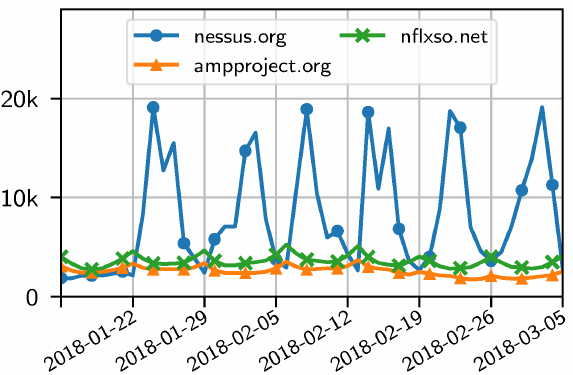}
		\vspace{-5mm}       
    \caption{Weekday/weekend dynamics in Umbrella \topm SLDs.}
		\label{subfig:slds_umbrella}
	\end{subfigure}
	\hfill
	\vspace{-3mm}
\caption{Comparison of weekday vs. weekend distributions and dynamics in Second-Level-Domains (SLDs).}
\vspace{-1em}
\end{figure*} 
\textbf{For How Long are Domains Part of a Top List?}
We investigate the average number of days a domain remains in both the \topm 
and \topk lists in \Cref{subfig:ndays}. 
This figure displays a CDF with the number of days from the JOINT dataset in 
the x-axis, and the normalised cumulative probability that a domain is included on the list for X or fewer days. 
Our analysis reveals significant differences across lists.  
While about 90\% of domains in the Alexa \topm list are in the list for 50 or fewer days,
40\% of domains in the Majestic \topm list remain in the list across the full year.
With this reading, lines closer to the lower right corner are ``better'' in the sense that more
domains have stayed in the list for longer periods, while lines closer to the upper left 
indicate that domains get removed more rapidly. 
The lists show quite different behaviour, with Majestic \topk being the most stable 
by far (only $\approx26\%$ of domains present on $<100\%$ of days), and being followed by
Majestic \topm, Umbrella \topk, Alexa \topk, Umbrella \topm, and Alexa \topm. 
The Majestic \topm list offers stability similar to the Alexa and Umbrella \topk lists.

\vspace{-.5em}
\subsection{Weekly Patterns}
\label{sec:weekly}

We now investigate the weekly\footnote{It is unclear what cut-off times list providers use, and how they offset time zones. For our analysis, we map files to days using our download timestamp} pattern in the Alexa and Umbrella lists as observed in \Cref{subfig:dailychange}. 
We generally do not include Majestic as it does not display a weekly pattern. 
In this section, we resort to various statistical methods to investigate those weekend patterns. 
We will describe each one of them in their relevant subsection.

\textbf{How Do Domain Ranks Change over the Weekends?}
The weekly periodical patterns shown in  \Cref{subfig:dailychange} show that list content depends on the day of the week. 
To investigate this pattern statistically, we calculate a weekday and weekend distribution of the rank
position of a given domain and compute the distance between those two
distribution using the Kolmogorov-Smirnov (KS) test.
This method allows us to statistically determine to what degree the distribution of a domain's ranks 
on weekdays and weekends overlap, and is shown in~\Cref{subfig:cdf_distance}. 
We include Majestic as a base line without a weekly pattern. 
For Alexa \topm, we can see that $\approx$$35$\% of domains have a KS distance of one, meaning that their weekend and weekday distributions have no data point in common.
This feature is also present in Umbrella's rank, where over 15\% of domains have a KS distance of 1. 
The changes are less pronounced for the \topk Alexa and Umbrella lists, suggesting that the top domains are more stable. 
As a reference, the KS distance when comparing weekdays to weekdays and
weekends to weekends is much lower. For 90\% of domains in Alexa or Umbrella
(\topk{} or \topm{}) the distance is lower than 0.05. The KS distance is lower
than 0.02 for all of the domains in Majestic rankings (\topk{} or \topm{}).
This demonstrates that a certain set of domains, the majority of them
localised in the long-tail, present disjunct rankings between weekends
and weekdays.

\textbf{What Domains are More Popular on Weekends?}
This leads to the question about the nature of domains changing in popularity
with a weekly pattern. 
To investigate this, we group domains by ``second-level-domain''
(SLD), which we define as the label left of a public suffix per the Public
Suffix list~\cite{pslgithub}. Figures~\ref{subfig:slds_alexa}
and~\ref{subfig:slds_umbrella} display the time dynamics of SLD groups for
which the number of domains varies by more than $40\%$ between weekdays and
weekends.
For Alexa, we can see stable behaviour before its February 2018 change.
We see that some groups such as \texttt{blogspot.*}\footnote{We include all
\texttt{blogspot.*} domains in the same group} or \texttt{tumblr.com} are significantly more popular on weekends than on weekdays.
The opposite is true for domains under \texttt{sharepoint.com} (a web-based Microsoft Office platform). 
Umbrella shows the same behaviour, with \texttt{nessus.org} (a threat intelligence tool) more popular during the week, and \texttt{ampproject.org} (a dominant website performance optimisation framework), and \texttt{nflxso.net} (a Netflix domain) more popular on weekends. 
These examples confirm that different Internet usage on weekends\footnote{Our data indicates prevailing Saturday and Sunday weekends} is a cause for the weekly patterns.

\subsection{Order of Domains in Top Lists}
As \toplists are sorted, a statistical analysis of order variation completes our view of top lists' stability.
We use the Kendall rank correlation coefficient~\cite{kendall1938new},
commonly known as Kendall's $\tau$ coefficient, to measure rank correlation,
\ie the similarity in the \textit{order} of lists.
Kendall's correlation between two variables will be high when observations have a similar
order between the two variables, and low when observations have a 
dissimilar (or fully different for a correlation of -1) rank between the two variables.

\begin{figure}	\centering
	\includegraphics[width=\columnwidth]{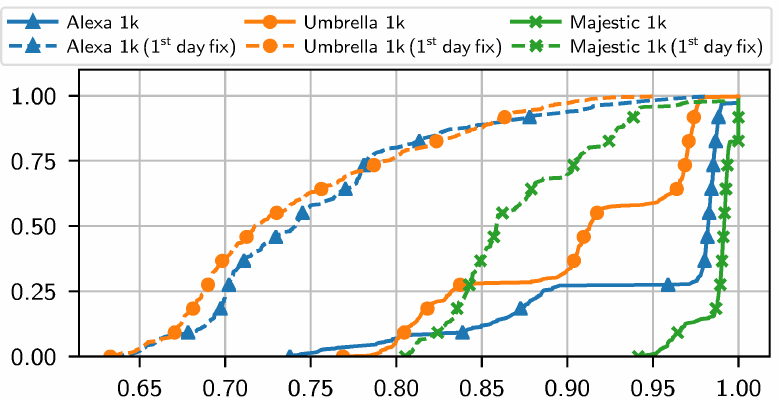}
	\vspace{-2em}
	\caption{CDF of Kendall's $\tau$ between top lists.}
	\label{subfig:tau}
	\vspace{-1em}
\end{figure}

In Figure~\ref{subfig:tau}, we show the CDF of Kendall's $\tau$ rank
correlation coefficient for the Alexa, Umbrella and Majestic \topk domains in
two cases:
\one for day to day comparisons; 
\two for a static comparison to the first day in the JOINT dataset.
For analysis, we can compare the percentage of very strongly correlated ranks,
\ie{} the ranks for which Kendall's $\tau$ is higher than 0.95. For day to day
comparisons, Majestic is clearly most similar at 99\%, with Alexa (72\%) and
Umbrella (40\%) both showing considerably dissimilarities.

\begin{table*}[ht]
\small
\centering
\caption{Rank variation for some more and less popular websites in the \topm lists.} 
\vspace{-3mm}
  \begin{tabular}{l rrr rrr rrr}
    \toprule
    \multirow{2}{*}{Domain} & \multicolumn{3}{c}{Highest rank} & \multicolumn{3}{c}{Median rank} & \multicolumn{3}{c}{Lowest rank} \\    
    \cmidrule(r){2-4}\cmidrule(r){5-7}\cmidrule(r){8-10}
      & Alexa & Umbrella & Majestic & Alexa & Umbrella & Majestic & Alexa & Umbrella & Majestic \\
    \cmidrule(r){1-1} \cmidrule(r){2-4}\cmidrule(r){5-7}\cmidrule(r){8-10}
    \texttt{google.com} & 1 & 1 & 1 & 1 & 1 & 1 & 2 & 4 & 8 \\
    \texttt{facebook.com} & 3 & 1 & 2 & 3 & 6 & 2 & 3 & 8 & 19 \\
    \texttt{netflix.com} & 21 & 1 & 455 & 32 & 2 & 515 & 34 & 487 & 572 \\
    \cmidrule(r){1-1} \cmidrule(r){2-4}\cmidrule(r){5-7}\cmidrule(r){8-10}    
    \texttt{jetblue.com} & 2,284 & 14,291 & 4,810 & 3,133 & 29,637 & 4,960 & 5,000 & 56,964 & 5,150 \\
    \texttt{mdc.edu} & 25,619 & 177,571 & 24,720 & 35,405 & 275,579 & 26,122 & 88,093 & 449,309 & 30,914 \\
    \texttt{puresight.com} & 183,088 & 593,773 & 687,838 & 511,800 & 885,269 & 749,819 & 998,407 & 999,694 & 869,872 \\
    \bottomrule& &   \end{tabular}
  \label{tab:rank_variation_exs}
\vspace{-2em}  
\end{table*}

When compared for a reference day, very strong correlation drops below 5\% for
all lists.
This suggests that the order variations are not perceived
in the short term, but may arise when considering longer temporal windows. 

\textbf{Investigating the Long Tail: } 
To compare higher and lower ranked domains, we take three exemplary 
domains from the Top 100 and the lower ranks as examples. 
Table~\ref{tab:rank_variation_exs} summarises the results. 
For each of the six domains, we compute the highest, median, and lowest rank
over the duration of the JOINT dataset.
The difference of variability between top and bottom domains is striking and in line with our previous findings: 
the ranks of top domains are fairly stable, while the ranks of bottom domains vary drastically.

\subsection{Summary}
We investigate the stability of top lists, and find abrupt changes, weekly patterns, and significant churn for some lists.
Lower ranked domains fluctuate more, but the effect heavily depends on the list and the subset (\topk or \topm).
We can confirm that the weekly pattern stems from leisure-oriented domains being more popular on weekends, 
and give examples for domain rank variations.

 \section{Understanding and Influencing Top Lists Ranking Mechanisms}\label{sec:influencing}
We have seen that top lists can be rather unstable from day to day, and hence we investigate what traffic levels are required and at what effort it is possible to manipulate the ranking of a certain domain.
As discussed previously, the Alexa list is based on its browser toolbar and ``various other sources'', Umbrella is based on OpenDNS queries, and Majestic is based on the count of subnets with inbound links to the domain in question. 
In this section, we investigate the ranking mechanisms of these top lists more closely.

\subsection{Alexa}
Alexa obtains visited URLs through ``over 25,000 different browser extensions'' to calculate site ranks through visitor and page view statistics~\cite{alexahowranking, alexamyths}.
There is no further information about these toolbars besides Alexa's own toolbar.
Alexa also provides data to The Internet Archive to add new sites~\cite{archivealexacrawl}.
It has been speculated that Alexa provides tracking information to feed the Amazon recommendation and profiling engine since Amazon's purchase of Alexa in 1999~\cite{alexaacquired}.
To better understand the ranking mechanism behind the Alexa list, we reverse engineer the Alexa toolbar\footnote{We detail the reverse engineering process in our dataset} and investigate what data it gathers.
Upon installation, the toolbar fetches a unique identifier which is stored in the browser's local storage, called the \textit{Alexa ID} (\textit{aid}). 
This identifier is used for distinctly tracking the device.
During installation, Alexa requests information about age, (binary) gender, household income, ethnicity, education, children, and the toolbar installation location (home/work).
All of these are linked to the \textit{aid}.
After installation, the toolbar transfers for each visited site: 
the page URL, screen/page sizes, referer, window IDs, tab IDs, and loading time metrics. 
For a scarce set of 8 search engine and shopping URLs\footnote{As of 2018-05-17, these are \texttt{google.com}, \texttt{instacart.com}, \texttt{shop.rewe.de}, \texttt{youtube.com}, \texttt{search.yahoo.com}, \texttt{jet.com} and \texttt{ocado.com}}, referer and URL are anonymised to their host name. 
For all other domains, the entire URL, including all GET parameters, is transmitted to Alexa's servers under \textit{data.alexa.com}.
Because of the injected JavaScript, the visit is only transmitted if the site actually exists and was loaded.
In April 2018, Alexa's API DNS name had a rank of $\approx$30k in the Umbrella list, indicating at least 10k unique source IP addresses querying that DNS domain name through OpenDNS per day (cf \Cref{subsec:influencing_umbrella}).

Due to its dominance, the Alexa rank of a domain is an important criterion in domain trading and search engine optimisation.
Unsurprisingly, there is a gray area industry of sites promising to ``optimise'' the Alexa rank of a site for money~\cite{alexaboostspecialist,alexaboostrankboostup,alexaboostupmyrank}.
Although sending synthetic data to Alexa's backend API should be possible at reasonable effort, we refrain from doing so for two reasons: \one in April 2018, the backend API has changed, breaking communication with the toolbar, and \two unclear ethical implications of actively injecting values into this API.
We refer the interested reader to le\,Pochat \etal~\cite{pochat2018rigging}, who have recently succeeded in manipulating Alexa ranks through the toolbar API.

\subsection{Umbrella}\label{subsec:influencing_umbrella}
As the Umbrella list is solely based on DNS queries through the OpenDNS public resolver, it mainly reflects domains frequently resolved, not necessarily domains visited by humans, as confirmed in \Cref{subsec:discrepancies}.
Examples are the Internet scanning machines of various research institutions, which likely show up in the Umbrella ranking through automated forward-confirmed reverse-DNS at scanned hosts, and not from humans entering the URL into their browser.
Building a top list based on DNS queries has various trade-offs and parameters, which we aim to explore here.
One specifically is the TTL value of a DNS domain name.
As the DNS highly relies on caching, TTL values could introduce a bias in determining popularity based on DNS query volume:
domain names with higher Time-To-Live values can be cached longer and may cause fewer DNS queries at upstream resolvers.
To better understand Umbrella's ranking mechanism and query volume required, 
we set up 7 RIPE Atlas measurements~\cite{RA1}, which query the OpenDNS resolvers for  DNS names under our control.

\textbf{Probe Count versus Query Volume: }\textsc{\begin{figure}
		\centering
		\begin{subfigure}[t]{.4975\columnwidth}
		\includegraphics[width=1.\columnwidth]{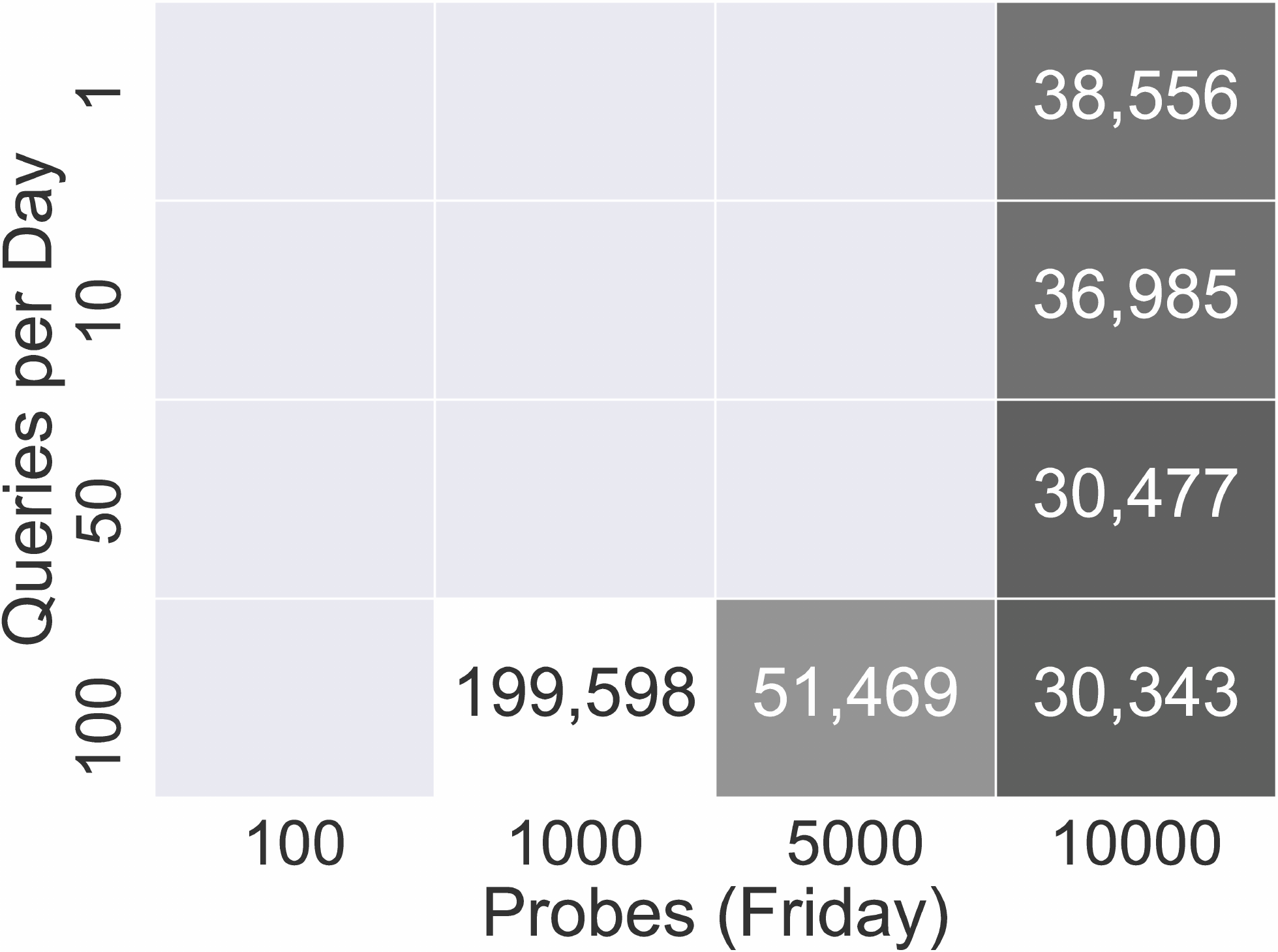}
		\end{subfigure}
		\begin{subfigure}[t]{.4925\columnwidth}
		\includegraphics[width=1.\columnwidth]{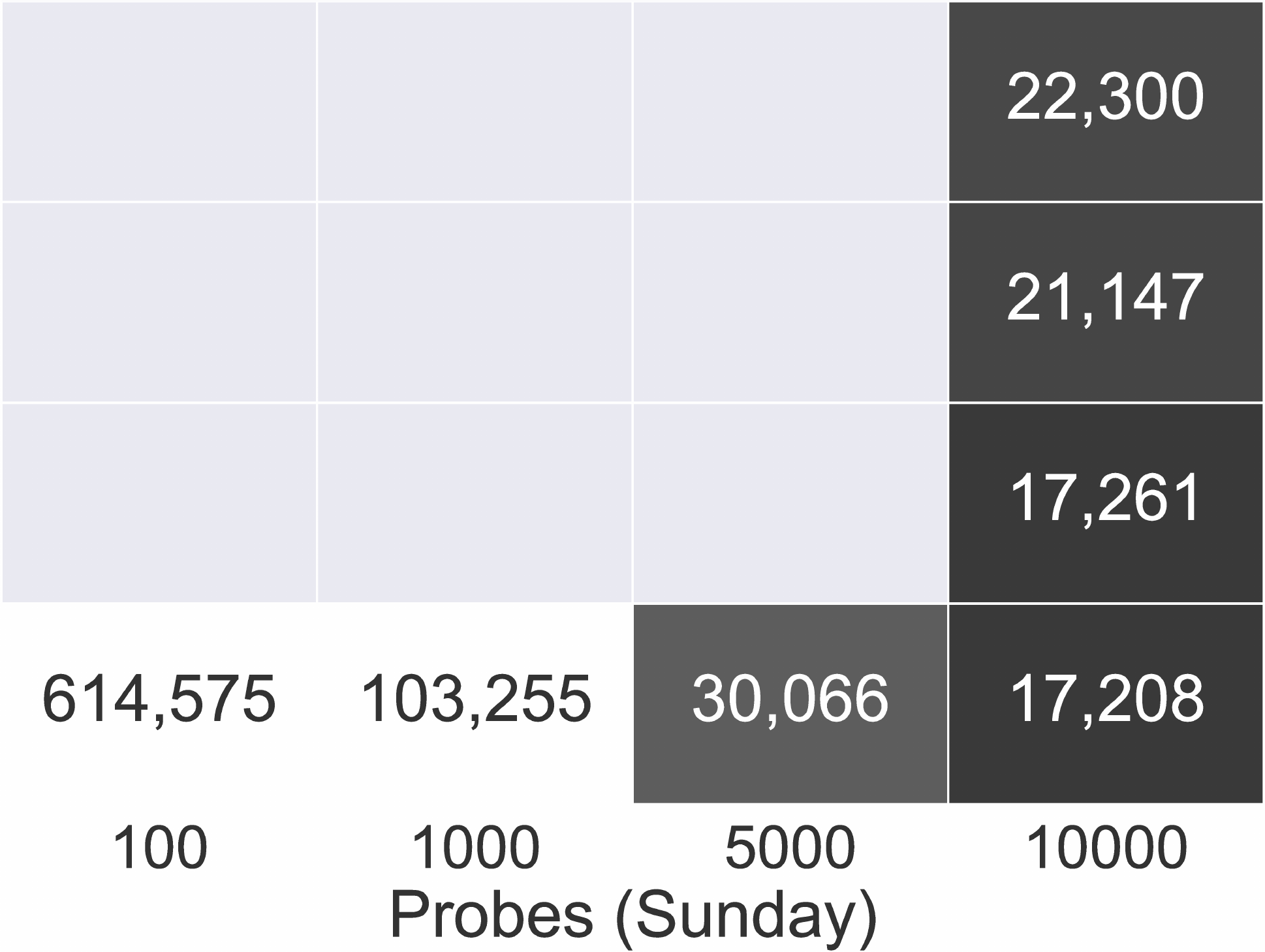}
		\end{subfigure}	
		\vspace{-2em}
		\caption{Umbrella rank depending on probe count, query frequency, and weekday (Friday left, Sunday right).
			Empty fields indicate the settings did not result in a \topm ranking.}		\label{fig:umbrellahacking}
\end{figure}}We set up measurements with 100, 1k, 5k, and 10k RIPE Atlas probes, and at frequencies of 1, 10, 50, and 100 DNS queries per RIPE Atlas probe per day~\cite{RA1}.
The resulting ranks, stabilised after several days of measurement, are depicted in \Cref{fig:umbrellahacking}.
A main insight is that the number of probes has a much stronger influence than the query volume per probe: 
10k probes at 1 query per day (a total of 10k queries) achieve a rank of 38k, while 1000 probes at 100 queries per day (a total of 100k queries) only achieve rank 199k.

It is a reasonable and considerate choice to base the ranking mechanism mainly on the number of unique sources, as it makes the ranking less susceptible to individual heavy hitters.

Upon stopping our measurements, our test domains quickly (within 1-2 days) disappeared from the list.

\textbf{TTL Influence: }
To test whether the Umbrella list normalises the potential effects of TTL values, we query DNS names with 5 different TTL values from 1000 probes at a 900s interval~\cite{RA2}.
We could not determine any significant effect of the TTL values: all 5 domains maintain a distance of less than 1k list places over time.

This is coherent with our previous observation that the Umbrella rank is mainly determined from the number of clients and not the query volume per client:
as the TTL volume would mainly impact the query volume per client, its effect should be marginal.

\subsection{Majestic}
The Majestic Million \toplist is based on a custom web crawler mainly used for commercial link intelligence~\cite{majesticabout}.
Initially, Majestic ranked sites by the raw number of referring domains.
As this had an undesired outcome, the link count was normalised by the count of referring \texttt{/24}-IPv4-subnets to limit the influence of single IP addresses~\cite{majesticnewalgorithm}.
The list is calculated using 90 days of data~\cite{majesticinsights}.
As this approach is similar to PageRank~\cite{pagerank}, except that Majestic does not weigh incoming links by the originating domain, it is to be expected that referral services can increase a domain's popularity.
We can, however, not see an \textit{efficient} way to influence a domain's rank in the Majestic list without using such referral services.
Le\,Pochat \etal \cite{pochat2018rigging} recently influenced a domain's rank in the Majestic link through such purchasing of back links.

\def\AlexatotalHR{{1M}}
\def\AlexatotalHROneK{{1K}}

\def\AlexaHTTPTwoPCT{{26.6}}
\def\AlexaHTTPTwoPCTSTD{{0.88}}

\def\AlexaHTTPTwoOneKPCT{{47.5}}
\def\AlexaHTTPTwoOneKPCTSTD{{0.75}}

\def\AlexaCDNOneKPCT{{27.5}}
\def\AlexaCDNOneKPCTSTD{{0.89}}

\def\AlexaCDNPCT{{6}}
\def\AlexaCDNPCTSTD{{0.6}}

\def\AlexaCNAMEOneKPCT{{53.1}}
\def\AlexaCNAMEOneKPCTSTD{{1.1}}

\def\AlexaCNAMEPCT{{44.1}}
\def\AlexaCNAMEPCTSTD{{1}} \def\UmbrellatotalHR{{1M}}
\def\UmbrellatotalHROneK{{1K}}

\def\UmbrellaHTTPTwoPCT{{19.11}}
\def\UmbrellaHTTPTwoPCTSTD{{0.63}}

\def\UmbrellaHTTPTwoOneKPCT{{36.3}}
\def\UmbrellaHTTPTwoOneKPCTSTD{{2.4}}

\def\UmbrellaCDNOneKPCT{{29.9}}
\def\UmbrellaCDNOneKPCTSTD{{0.37}}

\def\UmbrellaCDNPCT{{10.14}}
\def\UmbrellaCDNPCTSTD{{0.63}}

\def\UmbrellaCNAMEOneKPCT{{44.46}}
\def\UmbrellaCNAMEOneKPCTSTD{{0.43}}

\def\UmbrellaCNAMEPCT{{27.86}}
\def\UmbrellaCNAMEPCTSTD{{1}}

 \def\MajestictotalHR{{1M}}
\def\MajestictotalHROneK{{1K}}

\def\MajesticHTTPTwoPCT{{19.8}}
\def\MajesticHTTPTwoPCTSTD{{0.15}}

\def\MajesticHTTPTwoOneKPCT{{36.6}}
\def\MajesticHTTPTwoOneKPCTSTD{{0.72}}

\def\MajesticCDNOneKPCT{{36.1}}
\def\MajesticCDNOneKPCTSTD{{0.22}}

\def\MajesticCDNPCT{{2.6}}
\def\MajesticCDNPCTSTD{{0.01}}

\def\MajesticCNAMEOneKPCT{{64.8}}
\def\MajesticCNAMEOneKPCTSTD{{0.34}}

\def\MajesticCNAMEPCT{{39.81}}
\def\MajesticCNAMEPCTSTD{{0.15}} \def\ComnetorgtotalHR{{157.24M}}
\def\ComnetorgtotalSTDHR{{172K}}

\def\ComnetorgHTTPTwoPCT{{7.84}}
\def\ComnetorgHTTPTwoPCTSTD{{0.08}}

\def\ComnetorgCDNPCT{{1.3}}
\def\ComnetorgCDNPCTSTD{{0.004}}

\def\ComnetorgCNAMEPCT{{51.4}}
\def\ComnetorgCNAMEPCTSTD{{1.7}}

\def\AlexaTopMTLSSuccessRate{{74.65}}

\def\AlexaTopKTLSSusccessRate{{89.6}}

\def\UmbrellaTopMTLSSuccessRate{{43.05}}

\def\UmbrellaTopKTLSSuccessRate{{66.2}}

\def\MajesticTopMTLSSuccessRate{{62.89}}

\def\MajesticTopKTLSSuccessRate{{84.7}}

\def\ComNetOrgTopMTLSSuccessRate{{36.69}}

\def\AlexaTopMHSTS{{12.17}}

\def\AlexaTopKHSTS{{22.9}}

\def\MajesticTopMHSTS{{8.44}}

\def\MajesticTopKHSTS{{27.4}}

\def\UmbrellaTopMHSTS{{11.65}}

\def\UmbrellaTopKHSTS{{13.0}}

\def\ComNetOrgHSTS{{7.63}}
 \def\AlexaTopFirstCdnName{Google}
\def\AlexaTopFirstCdnSharePCT{44.85}
\def\AlexaTopSecondCdnName{Akamai}
\def\AlexaTopSecondCdnSharePCT{18.77}
\def\AlexaTopThirdCdnName{WordPress}
\def\AlexaTopThirdCdnSharePCT{10.17}
\def\AlexaTopFourthCdnName{Fastly}
\def\AlexaTopFourthCdnSharePCT{8.15}
\def\AlexaTopFifthCdnName{Incapsula}
\def\AlexaTopFifthCdnSharePCT{6.69}

\def\UmbrellaTopFirstCdnName{Akamai}
\def\UmbrellaTopFirstCdnSharePCT{44.75}
\def\UmbrellaTopSecondCdnName{Amazon CloudFront}
\def\UmbrellaTopSecondCdnSharePCT{12.80}
\def\UmbrellaTopThirdCdnName{Google}
\def\UmbrellaTopThirdCdnSharePCT{10.84}
\def\UmbrellaTopFourthCdnName{Fastly}
\def\UmbrellaTopFourthCdnSharePCT{7.45}
\def\UmbrellaTopFifthCdnName{Instart Logic}
\def\UmbrellaTopFifthCdnSharePCT{6.83}

\def\MajesticTopFirstCdnName{Akamai}
\def\MajesticTopFirstCdnSharePCT{39.89}
\def\MajesticTopSecondCdnName{Incapsula}
\def\MajesticTopSecondCdnSharePCT{14.10}
\def\MajesticTopThirdCdnName{Amazon CloudFront}
\def\MajesticTopThirdCdnSharePCT{11.98}
\def\MajesticTopFourthCdnName{Fastly}
\def\MajesticTopFourthCdnSharePCT{10.81}
\def\MajesticTopFifthCdnName{Google}
\def\MajesticTopFifthCdnSharePCT{9.56}

\def\ComnetorgTopFirstCdnName{Google}
\def\ComnetorgFirstCdnSharePCT{71.17}
\def\ComnetorgTopSecondCdnName{Akamai}
\def\ComnetorgSecondCdnSharePCT{8.14}
\def\ComnetorgTopThirdCdnName{Fastly}
\def\ComnetorgThirdCdnSharePCT{6.04}
\def\ComnetorgTopFourthCdnName{Incapsula}
\def\ComnetorgFourthCdnSharePCT{4.95}
\def\ComnetorgTopFifthCdnName{Zenedge}
\def\ComnetorgFifthCdnSharePCT{4.29}

\def\AlexaOneKTopFirstCdnName{Akamai}
\def\AlexaOneKTTopFirstCdnSharePCT{57.84}
\def\AlexaOneKTopSecondCdnName{Fastly}
\def\AlexaOneKTTopSecondCdnSharePCT{14.56}
\def\AlexaOneKTopThirdCdnName{Amazon CloudFront}
\def\AlexaOneKTTopThirdCdnSharePCT{10.41}
\def\AlexaOneKTopFourthCdnName{Google}
\def\AlexaOneKTTopFourthCdnSharePCT{4.00}
\def\AlexaOneKTopFifthCdnName{ChinaNetCenter}
\def\AlexaOneKTTopFifthCdnSharePCT{2.21}

\def\UmbrellaOneKTopFirstCdnName{Akamai}
\def\UmbrellaOneKTTopFirstCdnSharePCT{55.09}
\def\UmbrellaOneKTopSecondCdnName{Google}
\def\UmbrellaOneKTTopSecondCdnSharePCT{25.21}
\def\UmbrellaOneKTopThirdCdnName{Facebook}
\def\UmbrellaOneKTTopThirdCdnSharePCT{6.39}
\def\UmbrellaOneKTopFourthCdnName{Fastly}
\def\UmbrellaOneKTTopFourthCdnSharePCT{4.14}
\def\UmbrellaOneKTopFifthCdnName{Highwinds}
\def\UmbrellaOneKTTopFifthCdnSharePCT{1.65}

\def\MajesticOneKTopFirstCdnName{Akamai}
\def\MajesticOneKTTopFirstCdnSharePCT{52.91}
\def\MajesticOneKTopSecondCdnName{Fastly}
\def\MajesticOneKTTopSecondCdnSharePCT{19.66}
\def\MajesticOneKTopThirdCdnName{Google}
\def\MajesticOneKTTopThirdCdnSharePCT{9.46}
\def\MajesticOneKTopFourthCdnName{Amazon CloudFront}
\def\MajesticOneKTTopFourthCdnSharePCT{7.36}
\def\MajesticOneKTopFifthCdnName{Incapsula}
\def\MajesticOneKTTopFifthCdnSharePCT{2.21}

\def\AlexaMonTopFirstCdnName{Google}
\def\AlexaMonTopFirstCdnSharePCT{50.64}

\def\AlexaMonTopSecondCdnName{Akamai}
\def\AlexaMonTopSecondCdnSharePCT{15.67}

\def\AlexaMonTopThirdCdnName{WordPress}
\def\AlexaMonTopThirdCdnSharePCT{11.85}

\def\AlexaMonTopFourthCdnName{Fastly}
\def\AlexaMonTopFourthCdnSharePCT{7.15}

\def\AlexaMonTopFifthCdnName{Incapsula}
\def\AlexaMonTopFifthCdnSharePCT{5.43}

\def\AlexaTueTopFirstCdnName{Google}
\def\AlexaTueTopFirstCdnSharePCT{41.96}

\def\AlexaTueTopSecondCdnName{Akamai}
\def\AlexaTueTopSecondCdnSharePCT{20.27}

\def\AlexaTueTopThirdCdnName{WordPress}
\def\AlexaTueTopThirdCdnSharePCT{9.40}

\def\AlexaTueTopFourthCdnName{Fastly}
\def\AlexaTueTopFourthCdnSharePCT{8.66}

\def\AlexaTueTopFifthCdnName{Incapsula}
\def\AlexaTueTopFifthCdnSharePCT{7.31}

\def\AlexaWedTopFirstCdnName{Google}
\def\AlexaWedTopFirstCdnSharePCT{41.30}

\def\AlexaWedTopSecondCdnName{Akamai}
\def\AlexaWedTopSecondCdnSharePCT{20.49}

\def\AlexaWedTopThirdCdnName{WordPress}
\def\AlexaWedTopThirdCdnSharePCT{9.52}

\def\AlexaWedTopFourthCdnName{Fastly}
\def\AlexaWedTopFourthCdnSharePCT{8.80}

\def\AlexaWedTopFifthCdnName{Incapsula}
\def\AlexaWedTopFifthCdnSharePCT{7.36}

\def\AlexaThuTopFirstCdnName{Google}
\def\AlexaThuTopFirstCdnSharePCT{41.03}

\def\AlexaThuTopSecondCdnName{Akamai}
\def\AlexaThuTopSecondCdnSharePCT{20.60}

\def\AlexaThuTopThirdCdnName{WordPress}
\def\AlexaThuTopThirdCdnSharePCT{9.48}

\def\AlexaThuTopFourthCdnName{Fastly}
\def\AlexaThuTopFourthCdnSharePCT{8.84}

\def\AlexaThuTopFifthCdnName{Incapsula}
\def\AlexaThuTopFifthCdnSharePCT{7.45}

\def\AlexaFriTopFirstCdnName{Google}
\def\AlexaFriTopFirstCdnSharePCT{42.14}

\def\AlexaFriTopSecondCdnName{Akamai}
\def\AlexaFriTopSecondCdnSharePCT{20.20}

\def\AlexaFriTopThirdCdnName{WordPress}
\def\AlexaFriTopThirdCdnSharePCT{9.59}

\def\AlexaFriTopFourthCdnName{Fastly}
\def\AlexaFriTopFourthCdnSharePCT{8.55}

\def\AlexaFriTopFifthCdnName{Incapsula}
\def\AlexaFriTopFifthCdnSharePCT{7.25}

\def\AlexaSatTopFirstCdnName{Google}
\def\AlexaSatTopFirstCdnSharePCT{44.50}

\def\AlexaSatTopSecondCdnName{Akamai}
\def\AlexaSatTopSecondCdnSharePCT{19.06}

\def\AlexaSatTopThirdCdnName{WordPress}
\def\AlexaSatTopThirdCdnSharePCT{9.63}

\def\AlexaSatTopFourthCdnName{Fastly}
\def\AlexaSatTopFourthCdnSharePCT{8.23}

\def\AlexaSatTopFifthCdnName{Incapsula}
\def\AlexaSatTopFifthCdnSharePCT{6.84}

\def\AlexaSunTopFirstCdnName{Google}
\def\AlexaSunTopFirstCdnSharePCT{50.45}

\def\AlexaSunTopSecondCdnName{Akamai}
\def\AlexaSunTopSecondCdnSharePCT{16.16}

\def\AlexaSunTopThirdCdnName{WordPress}
\def\AlexaSunTopThirdCdnSharePCT{11.12}

\def\AlexaSunTopFourthCdnName{Fastly}
\def\AlexaSunTopFourthCdnSharePCT{7.15}

\def\AlexaSunTopFifthCdnName{Incapsula}
\def\AlexaSunTopFifthCdnSharePCT{5.58}

\def\UmbrellaMonTopFirstCdnName{Akamai}
\def\UmbrellaMonTopFirstCdnSharePCT{46.08}

\def\UmbrellaMonTopSecondCdnName{Amazon CloudFront}
\def\UmbrellaMonTopSecondCdnSharePCT{13.07}

\def\UmbrellaMonTopThirdCdnName{Google}
\def\UmbrellaMonTopThirdCdnSharePCT{9.85}

\def\UmbrellaMonTopFourthCdnName{Fastly}
\def\UmbrellaMonTopFourthCdnSharePCT{7.27}

\def\UmbrellaMonTopFifthCdnName{Instart Logic}
\def\UmbrellaMonTopFifthCdnSharePCT{4.85}

\def\UmbrellaTueTopFirstCdnName{Akamai}
\def\UmbrellaTueTopFirstCdnSharePCT{44.64}

\def\UmbrellaTueTopSecondCdnName{Amazon CloudFront}
\def\UmbrellaTueTopSecondCdnSharePCT{12.82}

\def\UmbrellaTueTopThirdCdnName{Google}
\def\UmbrellaTueTopThirdCdnSharePCT{10.96}

\def\UmbrellaTueTopFourthCdnName{Fastly}
\def\UmbrellaTueTopFourthCdnSharePCT{7.44}

\def\UmbrellaTueTopFifthCdnName{Instart Logic}
\def\UmbrellaTueTopFifthCdnSharePCT{6.82}

\def\UmbrellaWedTopFirstCdnName{Akamai}
\def\UmbrellaWedTopFirstCdnSharePCT{44.32}

\def\UmbrellaWedTopSecondCdnName{Amazon CloudFront}
\def\UmbrellaWedTopSecondCdnSharePCT{12.72}

\def\UmbrellaWedTopThirdCdnName{Google}
\def\UmbrellaWedTopThirdCdnSharePCT{11.05}

\def\UmbrellaWedTopFourthCdnName{Fastly}
\def\UmbrellaWedTopFourthCdnSharePCT{7.50}

\def\UmbrellaWedTopFifthCdnName{Instart Logic}
\def\UmbrellaWedTopFifthCdnSharePCT{7.44}

\def\UmbrellaThuTopFirstCdnName{Akamai}
\def\UmbrellaThuTopFirstCdnSharePCT{44.56}

\def\UmbrellaThuTopSecondCdnName{Amazon CloudFront}
\def\UmbrellaThuTopSecondCdnSharePCT{12.70}

\def\UmbrellaThuTopThirdCdnName{Google}
\def\UmbrellaThuTopThirdCdnSharePCT{11.08}

\def\UmbrellaThuTopFourthCdnName{Fastly}
\def\UmbrellaThuTopFourthCdnSharePCT{7.50}

\def\UmbrellaThuTopFifthCdnName{Instart Logic}
\def\UmbrellaThuTopFifthCdnSharePCT{7.20}

\def\UmbrellaFriTopFirstCdnName{Akamai}
\def\UmbrellaFriTopFirstCdnSharePCT{44.51}

\def\UmbrellaFriTopSecondCdnName{Amazon CloudFront}
\def\UmbrellaFriTopSecondCdnSharePCT{12.73}

\def\UmbrellaFriTopThirdCdnName{Google}
\def\UmbrellaFriTopThirdCdnSharePCT{11.01}

\def\UmbrellaFriTopFourthCdnName{Fastly}
\def\UmbrellaFriTopFourthCdnSharePCT{7.52}

\def\UmbrellaFriTopFifthCdnName{Instart Logic}
\def\UmbrellaFriTopFifthCdnSharePCT{7.26}

\def\UmbrellaSatTopFirstCdnName{Akamai}
\def\UmbrellaSatTopFirstCdnSharePCT{44.43}

\def\UmbrellaSatTopSecondCdnName{Amazon CloudFront}
\def\UmbrellaSatTopSecondCdnSharePCT{12.77}

\def\UmbrellaSatTopThirdCdnName{Google}
\def\UmbrellaSatTopThirdCdnSharePCT{10.97}

\def\UmbrellaSatTopFourthCdnName{Fastly}
\def\UmbrellaSatTopFourthCdnSharePCT{7.51}

\def\UmbrellaSatTopFifthCdnName{Instart Logic}
\def\UmbrellaSatTopFifthCdnSharePCT{7.35}

\def\UmbrellaSunTopFirstCdnName{Akamai}
\def\UmbrellaSunTopFirstCdnSharePCT{44.86}

\def\UmbrellaSunTopSecondCdnName{Amazon CloudFront}
\def\UmbrellaSunTopSecondCdnSharePCT{12.81}

\def\UmbrellaSunTopThirdCdnName{Google}
\def\UmbrellaSunTopThirdCdnSharePCT{10.82}

\def\UmbrellaSunTopFourthCdnName{Fastly}
\def\UmbrellaSunTopFourthCdnSharePCT{7.41}

\def\UmbrellaSunTopFifthCdnName{Instart Logic}
\def\UmbrellaSunTopFifthCdnSharePCT{6.69}

\def\MajesticMonTopFirstCdnName{Akamai}
\def\MajesticMonTopFirstCdnSharePCT{40.00}

\def\MajesticMonTopSecondCdnName{Incapsula}
\def\MajesticMonTopSecondCdnSharePCT{14.21}

\def\MajesticMonTopThirdCdnName{Amazon CloudFront}
\def\MajesticMonTopThirdCdnSharePCT{11.91}

\def\MajesticMonTopFourthCdnName{Fastly}
\def\MajesticMonTopFourthCdnSharePCT{10.49}

\def\MajesticMonTopFifthCdnName{Google}
\def\MajesticMonTopFifthCdnSharePCT{9.71}

\def\MajesticTueTopFirstCdnName{Akamai}
\def\MajesticTueTopFirstCdnSharePCT{39.87}

\def\MajesticTueTopSecondCdnName{Incapsula}
\def\MajesticTueTopSecondCdnSharePCT{14.09}

\def\MajesticTueTopThirdCdnName{Amazon CloudFront}
\def\MajesticTueTopThirdCdnSharePCT{12.00}

\def\MajesticTueTopFourthCdnName{Fastly}
\def\MajesticTueTopFourthCdnSharePCT{10.83}

\def\MajesticTueTopFifthCdnName{Google}
\def\MajesticTueTopFifthCdnSharePCT{9.55}

\def\MajesticWedTopFirstCdnName{Akamai}
\def\MajesticWedTopFirstCdnSharePCT{39.76}

\def\MajesticWedTopSecondCdnName{Incapsula}
\def\MajesticWedTopSecondCdnSharePCT{14.06}

\def\MajesticWedTopThirdCdnName{Amazon CloudFront}
\def\MajesticWedTopThirdCdnSharePCT{12.06}

\def\MajesticWedTopFourthCdnName{Fastly}
\def\MajesticWedTopFourthCdnSharePCT{10.97}

\def\MajesticWedTopFifthCdnName{Google}
\def\MajesticWedTopFifthCdnSharePCT{9.51}

\def\MajesticThuTopFirstCdnName{Akamai}
\def\MajesticThuTopFirstCdnSharePCT{39.91}

\def\MajesticThuTopSecondCdnName{Incapsula}
\def\MajesticThuTopSecondCdnSharePCT{14.12}

\def\MajesticThuTopThirdCdnName{Amazon CloudFront}
\def\MajesticThuTopThirdCdnSharePCT{11.99}

\def\MajesticThuTopFourthCdnName{Fastly}
\def\MajesticThuTopFourthCdnSharePCT{10.73}

\def\MajesticThuTopFifthCdnName{Google}
\def\MajesticThuTopFifthCdnSharePCT{9.63}

\def\MajesticFriTopFirstCdnName{Akamai}
\def\MajesticFriTopFirstCdnSharePCT{40.04}

\def\MajesticFriTopSecondCdnName{Incapsula}
\def\MajesticFriTopSecondCdnSharePCT{14.15}

\def\MajesticFriTopThirdCdnName{Amazon CloudFront}
\def\MajesticFriTopThirdCdnSharePCT{11.88}

\def\MajesticFriTopFourthCdnName{Fastly}
\def\MajesticFriTopFourthCdnSharePCT{10.64}

\def\MajesticFriTopFifthCdnName{Google}
\def\MajesticFriTopFifthCdnSharePCT{9.61}

\def\MajesticSatTopFirstCdnName{Akamai}
\def\MajesticSatTopFirstCdnSharePCT{39.91}

\def\MajesticSatTopSecondCdnName{Incapsula}
\def\MajesticSatTopSecondCdnSharePCT{14.12}

\def\MajesticSatTopThirdCdnName{Amazon CloudFront}
\def\MajesticSatTopThirdCdnSharePCT{11.96}

\def\MajesticSatTopFourthCdnName{Fastly}
\def\MajesticSatTopFourthCdnSharePCT{10.79}

\def\MajesticSatTopFifthCdnName{Google}
\def\MajesticSatTopFifthCdnSharePCT{9.57}

\def\MajesticSunTopFirstCdnName{Akamai}
\def\MajesticSunTopFirstCdnSharePCT{39.85}

\def\MajesticSunTopSecondCdnName{Incapsula}
\def\MajesticSunTopSecondCdnSharePCT{14.06}

\def\MajesticSunTopThirdCdnName{Amazon CloudFront}
\def\MajesticSunTopThirdCdnSharePCT{12.01}

\def\MajesticSunTopFourthCdnName{Fastly}
\def\MajesticSunTopFourthCdnSharePCT{10.88}

\def\MajesticSunTopFifthCdnName{Google}
\def\MajesticSunTopFifthCdnSharePCT{9.51} \def\AlexaOneKTopFirstASName{{Akamai}}
\def\AlexaOneKTopFirstASN{{20940}}
\def\AlexaOneKTopFirstASSharePCT{{15.23}}
\def\AlexaOneKTopSecondASName{{Cloudflare}}
\def\AlexaOneKTopSecondASN{{13335}}
\def\AlexaOneKTopSecondASSharePCT{{13.89}}
\def\AlexaOneKTopThirdASName{{Google}}
\def\AlexaOneKTopThirdASN{{15169}}
\def\AlexaOneKTopThirdASSharePCT{{9.80}}
\def\AlexaOneKTopFourthASName{{Amazon}}
\def\AlexaOneKTopFourthASN{{16509}}
\def\AlexaOneKTopFourthASSharePCT{{6.96}}
\def\AlexaOneKTopFifthASName{{Fastly}}
\def\AlexaOneKTopFifthASN{{54113}}
\def\AlexaOneKTopFifthASSharePCT{{5.75}}
\def\AlexaTopFirstASName{{Cloudflare}}
\def\AlexaTopFirstASN{{13335}}
\def\AlexaTopFirstASSharePCT{{10.26}}
\def\AlexaTopSecondASName{{Amazon}}
\def\AlexaTopSecondASN{{16509}}
\def\AlexaTopSecondASSharePCT{{4.42}}
\def\AlexaTopThirdASName{{Google}}
\def\AlexaTopThirdASN{{15169}}
\def\AlexaTopThirdASSharePCT{{4.39}}
\def\AlexaTopFourthASName{{OVH}}
\def\AlexaTopFourthASN{{16276}}
\def\AlexaTopFourthASSharePCT{{3.86}}
\def\AlexaTopFifthASName{{GoDaddy}}
\def\AlexaTopFifthASN{{26496}}
\def\AlexaTopFifthASSharePCT{{2.74}}

\def\UmbrellaOneKTopFirstASName{{Google}}
\def\UmbrellaOneKTopFirstASN{{15169}}
\def\UmbrellaOneKTopFirstASSharePCT{{18.56}}
\def\UmbrellaOneKTopSecondASName{{Amazon}}
\def\UmbrellaOneKTopSecondASN{{16509}}
\def\UmbrellaOneKTopSecondASSharePCT{{10.45}}
\def\UmbrellaOneKTopThirdASName{{Akamai}}
\def\UmbrellaOneKTopThirdASN{{20940}}
\def\UmbrellaOneKTopThirdASSharePCT{{10.24}}
\def\UmbrellaOneKTopFourthASName{{Microsoft}}
\def\UmbrellaOneKTopFourthASN{{8075}}
\def\UmbrellaOneKTopFourthASSharePCT{{7.25}}
\def\UmbrellaOneKTopFifthASName{{Amazon}}
\def\UmbrellaOneKTopFifthASN{{14618}}
\def\UmbrellaOneKTopFifthASSharePCT{{6.01}}
\def\UmbrellaTopFirstASName{{Amazon}}
\def\UmbrellaTopFirstASN{{16509}}
\def\UmbrellaTopFirstASSharePCT{{9.45}}
\def\UmbrellaTopSecondASName{{Cloudflare}}
\def\UmbrellaTopSecondASN{{13335}}
\def\UmbrellaTopSecondASSharePCT{{7.18}}
\def\UmbrellaTopThirdASName{{Amazon}}
\def\UmbrellaTopThirdASN{{14618}}
\def\UmbrellaTopThirdASSharePCT{{6.17}}
\def\UmbrellaTopFourthASName{{Google}}
\def\UmbrellaTopFourthASN{{15169}}
\def\UmbrellaTopFourthASSharePCT{{5.54}}
\def\UmbrellaTopFifthASName{{Akamai}}
\def\UmbrellaTopFifthASN{{20940}}
\def\UmbrellaTopFifthASSharePCT{{5.37}}

\def\MajesticOneKTopFirstASName{{Akamai}}
\def\MajesticOneKTopFirstASN{{20940}}
\def\MajesticOneKTopFirstASSharePCT{{17.89}}
\def\MajesticOneKTopSecondASName{{Fastly}}
\def\MajesticOneKTopSecondASN{{54113}}
\def\MajesticOneKTopSecondASSharePCT{{10.21}}
\def\MajesticOneKTopThirdASName{{Amazon}}
\def\MajesticOneKTopThirdASN{{16509}}
\def\MajesticOneKTopThirdASSharePCT{{8.27}}
\def\MajesticOneKTopFourthASName{{Cloudflare}}
\def\MajesticOneKTopFourthASN{{13335}}
\def\MajesticOneKTopFourthASSharePCT{{6.70}}
\def\MajesticOneKTopFifthASName{{Google}}
\def\MajesticOneKTopFifthASN{{15169}}
\def\MajesticOneKTopFifthASSharePCT{{6.20}}
\def\MajesticTopFirstASName{{Cloudflare}}
\def\MajesticTopFirstASN{{13335}}
\def\MajesticTopFirstASSharePCT{{9.22}}
\def\MajesticTopSecondASName{{GoDaddy}}
\def\MajesticTopSecondASN{{26496}}
\def\MajesticTopSecondASSharePCT{{4.45}}
\def\MajesticTopThirdASName{{Amazon}}
\def\MajesticTopThirdASN{{16509}}
\def\MajesticTopThirdASSharePCT{{3.35}}
\def\MajesticTopFourthASName{{OVH}}
\def\MajesticTopFourthASN{{16276}}
\def\MajesticTopFourthASSharePCT{{3.02}}
\def\MajesticTopFifthASName{{Amazon}}
\def\MajesticTopFifthASN{{14618}}
\def\MajesticTopFifthASSharePCT{{2.26}}

\def\ComnetorgTopFirstASName{{GoDaddy}}
\def\ComnetorgTopFirstASN{{26496}}
\def\ComnetorgTopFirstASSharePCT{{25.99}}
\def\ComnetorgTopSecondASName{{Amazon}}
\def\ComnetorgTopSecondASN{{14618}}
\def\ComnetorgTopSecondASSharePCT{{4.38}}
\def\ComnetorgTopThirdASName{{Amazon}}
\def\ComnetorgTopThirdASN{{16509}}
\def\ComnetorgTopThirdASSharePCT{{4.05}}
\def\ComnetorgTopFourthASName{{1\&1}}
\def\ComnetorgTopFourthASN{{8560}}
\def\ComnetorgTopFourthASSharePCT{{3.23}}
\def\ComnetorgTopFifthASName{{Confluence Networks}}
\def\ComnetorgTopFifthASN{{40034}}
\def\ComnetorgTopFifthASSharePCT{{2.65}}

\def\AlexaTopFiveASSharePCT{{25.68}}
\def\AlexaOneKTopFiveASSharePCT{{52.68}}
\def\AlexaTopFiveASShareSTD{{0.67}}
\def\AlexaOneKTopFiveASShareSTD{{1.74}}

\def\UmbrellaTopFiveASSharePCT{{33.95}}
\def\UmbrellaOneKTopFiveASSharePCT{{53.33}}
\def\UmbrellaTopFiveASShareSTD{{1.06}}
\def\UmbrellaOneKTopFiveASShareSTD{{1.75}}

\def\MajesticTopFiveASSharePCT{{22.29}}
\def\MajesticOneKTopFiveASSharePCT{{51.74}}
\def\MajesticTopFiveASShareSTD{{0.17}}
\def\MajesticOneKTopFiveASShareSTD{{1.73}}

\def\ComnetorgTopFiveASSharePCT{40.22}
\def\ComnetorgTopFiveASShareSTD{0.09}

\def\AlexaUniqueAS{{19511}}
\def\AlexaUniqueASSTD{{597}}
\def\AlexaOneKUniqueAS{{256}}
\def\AlexaOneKUniqueASSTD{{5}}

\def\UmbrellaUniqueAS{{16922}}
\def\UmbrellaUniqueASSTD{{584}}
\def\UmbrellaOneKUniqueAS{{132}}
\def\UmbrellaOneKUniqueASSTD{{4}}

\def\MajesticUniqueAS{{17418}}
\def\MajesticUniqueASSTD{{61}}
\def\MajesticOneKUniqueAS{{250}}
\def\MajesticOneKUniqueASSTD{{3}}

\def\ComnetorgUniqueAS{{34876}}
\def\ComnetorgUniqueASSTD{{53}}

\def\AlexavSIXUniqueAS{{1856}}
\def\AlexavSIXUniqueASSTD{{56}}
\def\AlexavSIXOneKUniqueAS{{44}}
\def\AlexavSIXOneKUniqueASSTD{{5}}

\def\UmbrellavSIXUniqueAS{{2591}}
\def\UmbrellavSIXUniqueASSTD{{157}}
\def\UmbrellavSIXOneKUniqueAS{{26}}
\def\UmbrellavSIXOneKUniqueASSTD{{2}}

\def\MajesticvSIXUniqueAS{{1236}}
\def\MajesticvSIXUniqueASSTD{{793}}
\def\MajesticvSIXOneKUniqueAS{{48}}
\def\MajesticvSIXOneKUniqueASSTD{{30}}

\def\ComnetorgvSIXUniqueAS{{3025}}
\def\ComnetorgvSIXUniqueASSTD{{9}}

\newcommand{\hih}{\greentriu\,} \newcommand{\hil}{\redtrid\,}  \newcommand{\his}{\bluensq\,}  

\begin{table*}[t]
	\caption{Internet measurement characteristics compared across top lists and general population, usually given as $\mu \pm \sigma$. For each cell, we highlight if it significantly (50\%\textsuperscript{6}) exceeds \hih or falls behind \hil the base value (1k / 1M, 1M /  \textit{com/net/org}),\,or~not\,\his. \\
	\textbf{In almost all cases (\hih and \hil), top lists significantly distort the characteristics of the general population.}}	
	\resizebox{\textwidth}{!}
	{
	\centering
	\begin{tabular}{l r r rrrrrr}
		\toprule
&  Alexa & Umbrella & Majestic & Alexa & Umbrella & Majestic & com/net/org \\

			Study				& \AlexatotalHROneK											& \UmbrellatotalHROneK												& \MajestictotalHROneK											& \AlexatotalHR{}											& \UmbrellatotalHR{}											& \MajestictotalHR{}										& \ComnetorgtotalHR{} $\pm$ \ComnetorgtotalSTDHR{}\\
		\cmidrule(lr){1-1}\cmidrule(lr){2-4}\cmidrule(lr){5-7}\cmidrule(lr){8-8}
		
		NXDOMAIN\textsuperscript{1}	& 	\hil $\sim$0.0\% $\pm$ 0.0\%	& 	\hil $\sim$0.0\% $\pm$ 0.0\% & 	\hil $\sim$0.0\% $\pm$ 0.0\%	&\hil 0.13\% $\pm$ 0.02 			& \hih 11.51\% $\pm$ 0.9 					& \hih 2.66\% $\pm$ 0.09 					& 0.8\% $\pm$ 0.02 \\
		IPv6-enabled\textsuperscript{2} & \hih 22.7\% $\pm$  0.6 & \hih	22.6\% $\pm$  1.0	& 	\hih 20.7\% $\pm$  0.4	&  \hih 12.9\% $\pm$ 0.9 			& \hih 14.8\% $\pm$ 0.8 					& \hih 10.8\% $\pm$ 0.2 					& 4.1\% $\pm$ 0.2 \\
		CAA-enabled\textsuperscript{1} 		& 	\hih 15.3\% $\pm$  0.9	& 	\hih	5.6\% $\pm$  0.3	& \hih 27.9\% $\pm$  0.3	& \hih 1.7\% $\pm$ 0.1 			& \hih 1.0\% $\pm$ 0.0 					& \hih 1.5\% $\pm$ 0.0  					& 0.1\% $\pm$ 0.0\\
		\midrule
		CNAMEs\textsuperscript{3}				& \his \AlexaCNAMEOneKPCT{}\% $\pm$ \AlexaCNAMEOneKPCTSTD{} 		& \hih \UmbrellaCNAMEOneKPCT{}\% $\pm$ \UmbrellaCNAMEOneKPCTSTD{} 		& \hih \MajesticCNAMEOneKPCT{}\% $\pm$ \MajesticCNAMEOneKPCTSTD{} 	& \his \AlexaCNAMEPCT{}\% $\pm$ \AlexaCNAMEPCTSTD{} 			& \hil \UmbrellaCNAMEPCT{}\% $\pm$ \UmbrellaCNAMEPCTSTD{}		& \hih \MajesticCNAMEPCT{}\% $\pm$ \MajesticCNAMEPCTSTD{}		& \ComnetorgCNAMEPCT{}\% $\pm$ \ComnetorgCNAMEPCTSTD{}\\
		CDNs (via CNAME)\textsuperscript{3}		&\hih \AlexaCDNOneKPCT{}\% $\pm$ \AlexaCDNOneKPCTSTD{} 			&\hih \UmbrellaCDNOneKPCT{}\% $\pm$ \UmbrellaCDNOneKPCTSTD{} 			&\hih \MajesticCDNOneKPCT{}\% $\pm$ \MajesticCDNOneKPCTSTD{} 		& \hih\AlexaCDNPCT{}\% $\pm$ \AlexaCDNPCTSTD{}					&\hih \UmbrellaCDNPCT{}\% $\pm$ \UmbrellaCDNPCTSTD{}				&\hih \MajesticCDNPCT{}\% $\pm$ \MajesticCDNPCTSTD{}			&\ComnetorgCDNPCT{}\% $\pm$ \ComnetorgCDNPCTSTD{}\\
		Unique AS IPv4 (avg.)\textsuperscript{3,4} & \AlexaOneKUniqueAS{} $\pm$ \AlexaOneKUniqueASSTD{} & \UmbrellaOneKUniqueAS{} $\pm$ \UmbrellaOneKUniqueASSTD{} & \MajesticOneKUniqueAS{} $\pm$ \MajesticOneKUniqueASSTD{} & \AlexaUniqueAS{} $\pm$ \AlexaUniqueASSTD{} & \UmbrellaUniqueAS{} $\pm$ \UmbrellaUniqueASSTD{} & \MajesticUniqueAS{} $\pm$ \MajesticUniqueASSTD{} & \ComnetorgUniqueAS{} $\pm$ \ComnetorgUniqueASSTD{}\\
		Unique AS IPv6 (avg.)\textsuperscript{3,4} & \AlexavSIXOneKUniqueAS{} $\pm$ \AlexavSIXOneKUniqueASSTD{} & \UmbrellavSIXOneKUniqueAS{} $\pm$ \UmbrellavSIXOneKUniqueASSTD{} & \MajesticvSIXOneKUniqueAS{} $\pm$ \MajesticvSIXOneKUniqueASSTD{} & \AlexavSIXUniqueAS{} $\pm$ \AlexavSIXUniqueASSTD{} & \UmbrellavSIXUniqueAS{} $\pm$ \UmbrellavSIXUniqueASSTD{} & \MajesticvSIXUniqueAS{} $\pm$ \MajesticvSIXUniqueASSTD{} & \ComnetorgvSIXUniqueAS{} $\pm$ \ComnetorgvSIXUniqueASSTD{}\\
		Top 5 AS (Share)\textsuperscript{3} & \hih \AlexaOneKTopFiveASSharePCT{}\% $\pm$ \AlexaOneKTopFiveASShareSTD{} & \hih \UmbrellaOneKTopFiveASSharePCT{}\% $\pm$ \UmbrellaOneKTopFiveASShareSTD{} & \hih \MajesticOneKTopFiveASSharePCT{}\% $\pm$ \MajesticOneKTopFiveASShareSTD{} & \hih \AlexaTopFiveASSharePCT{}\% $\pm$ \AlexaTopFiveASShareSTD{} & \his \UmbrellaTopFiveASSharePCT{}\% $\pm$ \UmbrellaTopFiveASShareSTD{} & \hih \MajesticTopFiveASSharePCT{}\% $\pm$ \MajesticTopFiveASShareSTD{} & \ComnetorgTopFiveASSharePCT{} $\pm$ \ComnetorgTopFiveASShareSTD{}\\
		\midrule
		TLS-capable\textsuperscript{5} & \hih\AlexaTopKTLSSusccessRate{}\% & \hih \UmbrellaTopKTLSSuccessRate{}\% & \hih\MajesticTopKTLSSuccessRate{}\% & \hih \AlexaTopMTLSSuccessRate{}\% & \his \UmbrellaTopMTLSSuccessRate{}\% & \hih \MajesticTopMTLSSuccessRate{}\% & \ComNetOrgTopMTLSSuccessRate{}\% \\
		HSTS-enabled HTTPS\textsuperscript{5} & \hih \AlexaTopKHSTS{}\% & \his \UmbrellaTopKHSTS{}\% & \hih \MajesticTopKHSTS{}\% &\hih  \AlexaTopMHSTS{}\% & \hih \UmbrellaTopMHSTS{}\% & \his \MajesticTopMHSTS{}\% & \ComNetOrgHSTS{}\% \\   
HTTP2\textsuperscript{3}					& \hih \AlexaHTTPTwoOneKPCT{}\% $\pm$ \AlexaHTTPTwoOneKPCTSTD{} 		&\hih \UmbrellaHTTPTwoOneKPCT{}\% $\pm$ \UmbrellaHTTPTwoOneKPCTSTD{}	&\hih \MajesticHTTPTwoOneKPCT{}\% $\pm$ \MajesticHTTPTwoOneKPCTSTD{}	&\hih \AlexaHTTPTwoPCT{}\% $\pm$ \AlexaHTTPTwoPCTSTD{}			&\hih \UmbrellaHTTPTwoPCT{}\% $\pm$ \UmbrellaHTTPTwoPCTSTD{}		&\hih \MajesticHTTPTwoPCT{}\% $\pm$ \MajesticHTTPTwoPCTSTD{}	& \ComnetorgHTTPTwoPCT{}\% $\pm$ \ComnetorgHTTPTwoPCTSTD{}\\
		\bottomrule
    \noalign{\vskip 1mm}
		\multicolumn{8}{p{\paperwidth}}{\normalsize 1: $\mu$ Apr, 2018~~~~ 2: $\mu$ of JOINT period (6.6.17--30.4.18)~~~~ 3: $\mu$ Apr, 2018 - 8. May, 2018~~~~ 4: no share, thus no \hil, \his, or \hih~~~ 5: Single day/list in May, 2018 ~~~ 6: For base values over 40\%, the test for significant deviation is 25\% and 5$\sigma$.}	\end{tabular}
	}
	\label{tab:h2_analysis_test}
\end{table*}

 \begin{figure*}
	\begin{subfigure}[t]{0.32\textwidth}
		\includegraphics[width=\columnwidth]{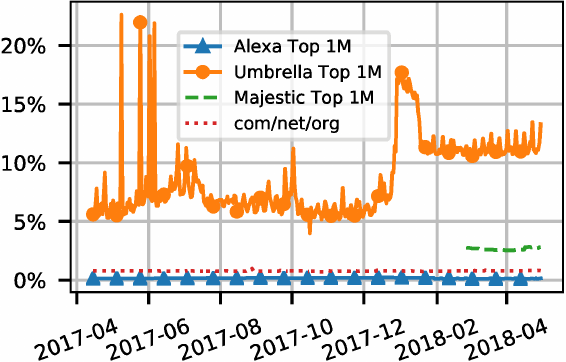} 		\vspace{-5mm}
		\caption{\% of NXDOMAIN responses.}
		\label{subfig:nxd}
	\end{subfigure}
	\hfill
	\begin{subfigure}[t]{0.32\textwidth}
		\includegraphics[width=\columnwidth]{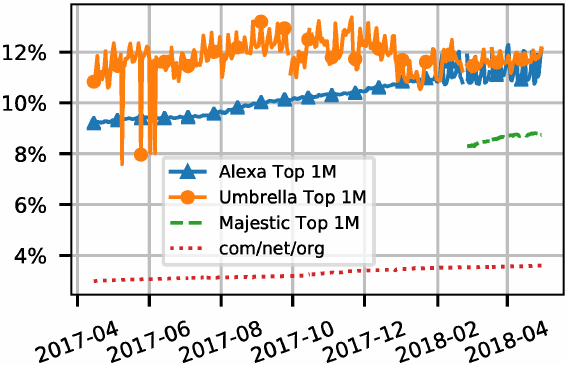} 		\vspace{-5mm}       
		\caption{\% of IPv6 Adoption.}
		\label{subfig:dns_ip6}
	\end{subfigure}
	\hfill
	\begin{subfigure}[t]{0.32\textwidth}
		\includegraphics{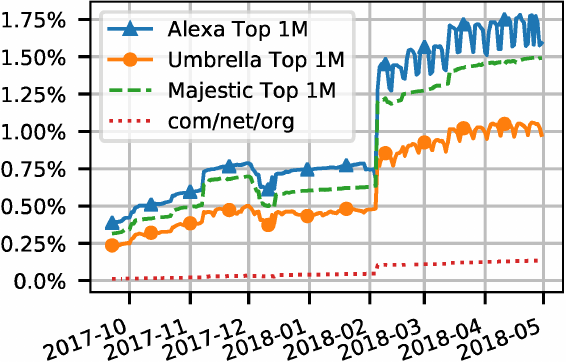} 		\vspace{-5mm}
		\caption{\% of CAA-enabled domains.}
		\label{subfig:dns_caa}
	\end{subfigure}
	\vspace{-3mm}
		\caption{DNS characteristics in the \topm lists and general population of about 158M domains.}
		\label{fig:dns}		
		\vspace{-3mm}
\end{figure*}

\section{Impact on Research Results}\label{sec:impact}
\Cref{sec:useoflists} highlighted that top lists are broadly used in networking, security and systems research.
Their use is especially prevalent in Internet measurement research, where top lists are used to study aspects across all layers.
This motivates us to understand the impact of top list usage on the outcome of these studies.
As the replication of all studies covered in our survey is not possible, we evaluate the impact of the lists' structure on research results in the Internet measurement field by investigating 
\first common layers, such as DNS and IP, that played a role in many studies, and 
 \second a sample of specific studies across a variety of layers, aiming for one specific study per layer. 

We evaluate those scientific results with 3 questions in mind:
\one what is the bias when using a top list as compared to a general population of all \textit{com/net/org} domains\footnote{\textit{com/net/org} is still only a 45\% sample of the general population (156.7M of 332M domains as per~\cite{dni}), but more complete and still unbiased samples are difficult to obtain due to ccTLDs' restrictive zone file access policies.~\cite{caastudy17, paper173:2017:imc, openintel_jsac, ipv6hitlist, h2adoption17}}
\two what is the difference in result when using a different top list?
\three what is the difference in result when using a top list from a different day?

\subsection{Domain Name System (DNS)}

A typical first step in list usage is DNS resolution, which is also a popular research focus (cf.~\Cref{sec:useoflists}).
We split this view into a record type perspective (\eg IPv6 adoption) and a hosting infrastructure perspective (\eg CDN prevalence and AS mapping).
For both, we download lists and run measurements daily over the course of one year. \subsubsection{Record Type Perspective}
We investigate the share of NXDOMAIN domains and IPv6-enabled domains, and the share of CAA-enabled domains as an example of a DNS-based measurement study~\cite{caastudy17}.
Results are shown in \Cref{tab:h2_analysis_test} and \Cref{fig:dns}.

\textbf{Assessing list quality via NXDOMAIN:}
We begin by using NXDOMAIN as a proxy measure for assessing the quality of entries in the top lists.
An NXDOMAIN error code in return to a DNS query means that the queried DNS name does not exist at the respective authoritative nameserver. This error code is unexpected for allegedly popular domains. Ideally, a top list would only provide existing domains.
Surprisingly, we find the amount of NXDOMAIN responses in both the Umbrella (11.5\%) and the Majestic (2.7\%) top lists higher than in the general population of \textit{com/net/org} domains (0.8\%). 
This is in alignment with the fact that already $\approx$23k domains in the Umbrella list belong to non-existent top-level domains (\cf \Cref{subsec:structure:tld}).
\Cref{subfig:nxd} shows that the NXDOMAIN share is, except for Umbrella, stable over time.
We found almost no NXDOMAINs among \topk ranked domains. One notable exception is \texttt{teredo.ipv6.microsoft.com}, a service discontinued in 2013 and unreachable, but still commonly appearing at high ranks in Umbrella, probably through requests from legacy clients.

This also highlights a challenge in Majestic's ranking mechanism: 
while counting the number of links to a certain website is quite stable over time, it also reacts slowly to domain closure.

\textbf{Tracking IPv6 adoption} has been the subject of several scientific studies such as~\cite{czyzv6, eravuchira2016measuring}.
We compare IPv6 adoption across \toplists and the general population, for which we count the number of domains that return at least one routed IPv6 address as an AAAA record or within a chain of up to 10 CNAMEs.
At 11--13\%, we find IPv6 enablement across top lists to significantly exceed the general population of domains at 4\%.
Also, the highest adoption lies with Umbrella, a good indication for IPv6 adoption: 
when the most frequently resolved DNS names support IPv6, many subsequent content requests are enabled to use IPv6.

\textbf{CAA Adoption:}
Exemplary for other record types, we also investigate the adoption of Certification Authority Authorization (CAA) records in top lists and the general population.  
CAA is a rather new record type, and has become mandatory for CAs to check before certificate issuance, \cf~\cite{caastudy17,ruohonen2018empirical}.
We measure CAA adoption as described in~\cite{caastudy17}, \ie the count of base domains with an \textit{issue} or \textit{issuewild} set.
Similar to IPv6 adoption, we find CAA adoption among top lists (1--2\%) to significantly exceed adoption among the general population at 0.1\%.
Even more stunning, the \topk lists feature a CAA adoption of up to 28\%, distorting the 0.1\% in the general population by two magnitudes. 

\textbf{Takeaway:} The DNS-focused results above highlight that \toplists may introduce a picture where results significantly differ from the general population, a popularity bias to be kept in mind. 
\Cref{fig:dns} also shows that Umbrella, and recently Alexa, can have different results when using a different day.
The daily differences, ranging, \eg from 1.5--1.8\% of CAA adoption around a mean of 1.7\% for Alexa, are not extreme, but should be accounted for.

\begin{figure*}
	\captionsetup[subfigure]{width=0.975\textwidth, justification=centering}
	\begin{subfigure}[t]{0.24\textwidth}
		\includegraphics[width=\columnwidth]{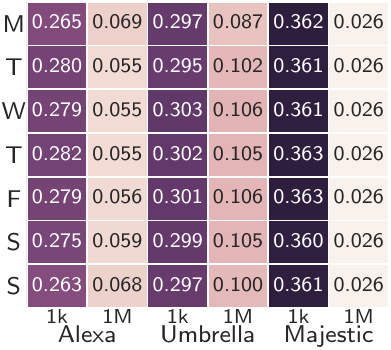}
		\vspace{-5mm}
		\caption{Ratio of detected CDNs by\\ list (x-axis) \& weekday (y-axis).}
		\label{subfig:cdns:share}
	\end{subfigure}	\begin{subfigure}[t]{0.24\textwidth}
		\includegraphics[width=\columnwidth]{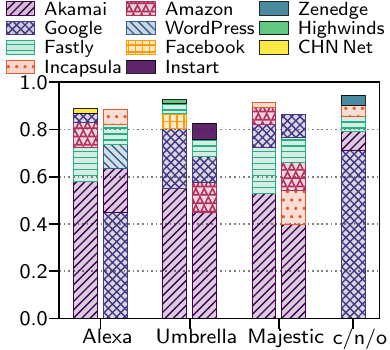}
		\vspace{-5mm}
		\caption{Share of top 5 CDNs,\\\topk vs. \topm vs. \textit{com/net/org}.}
		\label{subfig:cdns:top}
	\end{subfigure}	\begin{subfigure}[t]{0.24\textwidth}
		\includegraphics[width=\columnwidth]{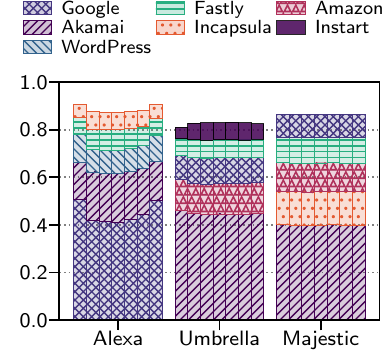}
		\vspace{-5mm}
		\caption{Share of top 5 CDNs,\\daily pattern (Mon - Sun).}
		\label{subfig:cdns:week}
	\end{subfigure}	\begin{subfigure}[t]{0.24\textwidth}
		\includegraphics[width=\columnwidth]{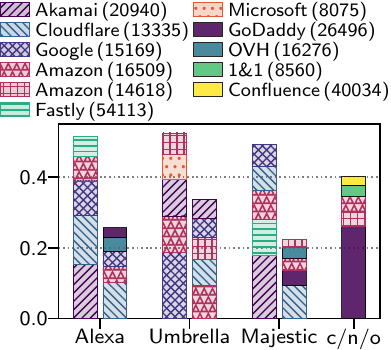}
		\vspace{-5mm}
		\caption{Share of top 5 ASes,\\\topk vs. \topm vs. \textit{com/net/org}}
		\label{subfig:asns}
	\end{subfigure}	\caption{Overall CDN ratio, ratio of top 5 CDNs, and ratio of top 5 ASes, dependent on list, list type, and weekday.}
	\label{fig:cdns}	
	\vspace{-3mm}
\end{figure*}

\subsubsection{Hosting Infrastructure Perspective}
Domains can be hosted by users themselves, by hosting companies, or a variety of CDNs.
The hosting landscape is subject to a body of research that is using \toplists to obtain target domains.
Here, we study the share of hosting infrastructures in different \toplists.

\textbf{CDN Prevalence:}\label{subsec:cdn}
We start by studying the prevalence of CDNs in \toplists and the general population of all \textit{com/net/org} domains. 
Since many CDNs use DNS CNAME records, we perform daily DNS resolutions in April 2018, querying all domains both raw \textit{www}-prefixed. We match the observed CNAME records against a list of CNAME patterns for 77 CDNs~\cite{wptcdndns} to identify CDN use. 

We first observe that the prevalence of CDNs differs by list and domain rank (see Table~\ref{tab:h2_analysis_test}), with all \topm lists exceeding the general population by at least a factor of 2, and all \topk lists exceeding the general population by at least a factor of 20.
When grouping the CDN ratio per list by weekdays (see Figure~\ref{subfig:cdns:share}), we observe minor influences of weekends vs.\ weekdays due to the top list dynamics described in ~\Cref{sec:weekly}.

After adoption of CDNs in general, we study the structure of CDN adoption.
We analyse the top 5 CDNs and show their distribution in Figure~\ref{fig:cdns} to study if the relative share is stable over different lists.
We thus show the fraction of domains using one of the top 5 CDNs for both a subset of the \topk and the entire list of \topm domains per list.
We first observe that the relative share of the top 5 CDNs differs by list and rank (see Figure~\ref{subfig:cdns:top}), but is generally very high at $>$80\%.
The biggest discrepancy is between using a top list and focusing on the general population of \textit{com/net/org} domains.
Google dominates the general population with a share of \ComnetorgFirstCdnSharePCT\% due to many (private) Google-hosted sites.
Domains in top lists are more frequently hosted by typical CDNs (\eg Akamai).
Grouping the CDN share per list by weekday in \Cref{subfig:cdns:week} shows a strong weekend/weekday pattern for Alexa, due to the rank dynamics discussed in~\Cref{sec:weekly}).
Interestingly, the weekend days have a higher share of Google DNS, indicating that more privately-hosted domains are visited on the weekend. 

These observations highlight that using a \toplist or not has significant influence on the top 5 CDNs observed, and, if using Alexa, the day of list creation as well.

\textbf{ASes:}\label{subsec:asn}
We next analyse the distribution of Autonomous Systems (AS) that announce a DNS name's A record in BGP, as per Route Views data from the day of the measurement, obtained from~\cite{routeviews}.
First, we study the AS diversity by counting the number of different ASes hit by the different lists. 
We observe lists to experience large differences in the number of unique ASes (\cf~\Cref{tab:h2_analysis_test}); while Alexa \topm hits the most ASes, \ie~\AlexaUniqueAS{} on average, Umbrella \topm hits the fewest, \ie~\UmbrellaUniqueAS{} on average. 
To better understand which ASes contribute the most IPs, we next focus on studying the top ASes.
Figure~\ref{subfig:asns} shows the top 5 ASes for the \topk and \topm domains of each list, as well as the set of \textit{com/net/org} domains.
We observe that both the set and share of involved ASes differ by list.

We note that the general share of the top 5 ASes is 40\% in the general population, compared to an average of 53\% in the \topk and an average of 27\% in the \topm lists.

In terms of structure, we further observe that  \ComnetorgTopFirstASName{} (AS\ComnetorgTopFirstASN{}) clearly dominates the general population with a share of \ComnetorgTopFirstASSharePCT{}\%, while it only accounts for \AlexaTopFifthASSharePCT{}\% on the Alexa \topm and for \MajesticTopSecondASSharePCT{}\% on the Majestic \topm.

While Alexa and Majestic share a somewhat similar distribution for both the \topm and \topk lists, Umbrella offers a quite different view, with a high share of Google/AWS hosted domains, which also relates to the CDN analysis above. 

This view is also eye-opening for other measurement studies: 
with a significant share of a population hosted by different 5 ASes, it is of no surprise that certain higher layer characteristics differ.

\subsection{TLS}
In line with the prevalence of TLS studies amongst the surveyed top list papers in \Cref{sec:useoflists}, we next investigate TLS adoption among lists and the general population. To probe for TLS support, we instruct \texttt{zgrab} to visit each domain via HTTPS for one day per list in May 2018.
As in the previous section, we measure all domains with and without \textit{www} prefix (except for Umbrella that contains subdomains), as we found greater coverage for these domains.
We were able to successfully establish TLS connections with \AlexaTopMTLSSuccessRate{{}}\% of the Alexa, \MajesticTopMTLSSuccessRate{{}}\% of the Majestic, \UmbrellaTopMTLSSuccessRate{{}}\% of the Umbrella, and \ComNetOrgTopMTLSSuccessRate{{}}\% of the \textit{com/net/org} domains (\cf Table~\ref{tab:h2_analysis_test}).
For \topk domains, TLS support further increases by 15--30\% per list.

These results show TLS support to be most pronounced among Alexa-listed domains, and that support in top lists generally exceeds the general population.

\textbf{HSTS:} As one current research topic~\cite{paper173:2017:imc}, we study the prevalence of HTTP Strict Transport Security (HSTS) among TLS enabled domains.
We define a domain to be HSTS--enabled if the domain provides a valid HSTS header with a \textit{max-age} setting $>$0. 
Out of the TLS-enabled domains, \AlexaTopMHSTS\% of the Alexa, \UmbrellaTopMHSTS\% of the Umbrella, \MajesticTopMHSTS\% of the Majestic, and \ComNetOrgHSTS\% of the com/net/org domains provide HSTS support (see Table~\ref{tab:h2_analysis_test}).
Only inspecting \topk domains again increases support significantly to \AlexaTopKHSTS\% for Alexa, \UmbrellaTopKHSTS\% for Umbrella, and \MajesticTopKHSTS\% for Majestic.
HSTS support is, again, over-represented in top lists.

\subsection{HTTP/2 Adoption}
One academic use of top lists is to study the adoption of upcoming protocols, \eg HTTP/2~\cite{webh2yet16,h2adoption17}.
The motivation for probing top listed domains can be based on the assumption that popular domains are more likely to adopt new protocols and are thus promising targets to study.
We thus exemplify this effect and the influence of different top lists by probing domains in top lists and the general population for their HTTP/2 adoption. 

We try to fetch the domains' landing page via HTTP/2 by using the \texttt{nghttp2} library.
We again \textit{www}-prefix all domains in Alexa and Majestic. 
In case of a successfully established HTTP/2 connection, we issue a \texttt{GET} request for the \textit{/} page of the domain.
We follow up to 10 redirects and if actual data for the landing page is transferred via HTTP/2, we count the domain as HTTP/2-enabled.
We probe top lists on a daily basis and the larger zone file on a weekly basis. 

\begin{figure}	\includegraphics[width=\columnwidth]{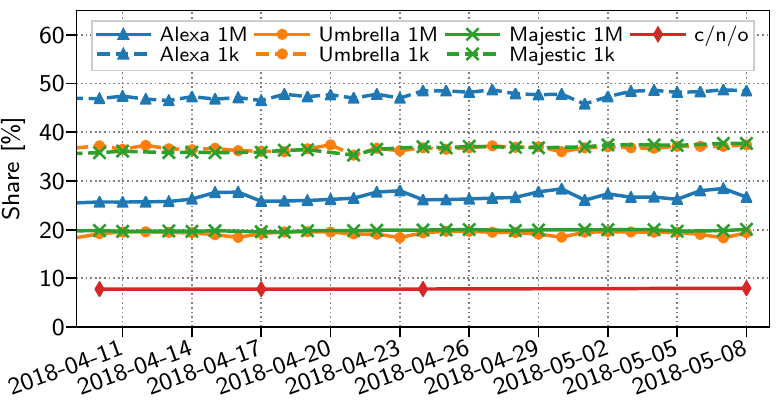}
	\vspace{-5mm}
	\caption{HTTP/2 adoption over time for the \topk and \topm lists and \textit{com/net/org} domains.}
	\label{fig:h2_adoption}
	\vspace{-5mm}
\end{figure}

We show HTTP/2 adoption in Figure~\ref{fig:h2_adoption}.
First, we observe that the HTTP/2 adoption of all \textit{com/net/org} domains is \ComnetorgHTTPTwoPCT{}\% on average and thus significantly lower than for domains listed in \topm lists, (up to \AlexaHTTPTwoPCT\% for Alexa) and even more so for \topk lists, which show adoption around 35\% or more.

One explanation is that, as shown above, popular domains are more likely hosted on progressive infrastructures (\eg CDNs) than the general population.

We next investigate HTTP/2 adoption between top lists based on Figure~\ref{fig:h2_adoption}.
Unsurprisingly, we observe HTTP/2 adoption differs by list and by weekday for those lists with a weekday pattern (\cf \Cref{sec:weekly}).
We also note the extremely different result when querying the \topk lists as compared to the general population.

\subsection{Takeaway} 
We have analysed the properties of top lists and the general population across many layers, and found that
top lists \one generally show significantly more extreme measurement results, \eg protocol adoption. This effect is pronounced to up to 2 orders of magnitude for the \topk domains.
Results can \two be affected by a weekly pattern, \eg the \% of protocol adoption may yield a different result when using a list generated on a weekend as compared to a weekday.
This is a significant limitation to be kept in mind when using top lists for measurement studies.

 \section{Discussion}\label{sec:discussion}
We have shown in \Cref{sec:significance} that \toplists are being frequently used in scientific studies. 
We acknowledge that using \toplists has distinct advantages---they provide a set of relevant domains at a small and stable size that can be compared over time.
However, the use of top lists also comes with certain disadvantages, which we have explored in this paper.

First, while it is the stated purpose of a \toplist to provide a sample biased towards the list's specific measure of popularity, these samples  do not represent the general state in the Internet well: 
we have observed in \Cref{sec:impact} that almost all conceivable measurements suffer significant bias when using a \topm list, and excessive bias in terms of magnitudes when using a \topk list. 
This indicates that domains in \toplists exhibit behaviour significantly different from the general population---quantitative insights based on top list domains likely will not generalise.

Second, we have shown that \toplists can significantly change from day to day, rendering results of one-off measurements unstable. A similar effect is that lists may be structurally different on weekends and weekdays, yielding differences in results purely based on the day of week when a list was downloaded. 

Third, the choice of a certain \toplist can significantly influence measurement results as well, \eg for CDN or AS structure (\cf \Cref{subsec:asn}), which stems from different lists having different sampling biases.
While these effects can be desired, \eg to find many domains that adopt a certain new technology, it leads to bad generalisation of results to ``the Internet'', and results obtained from measuring top lists must be interpreted very carefully. 

\subsection{Recommendation for Top List Use}
Based on our observations, we develop specific recommendations for the use of \toplists.
\Cref{sec:significance} has revealed that \toplists are used for different purposes in diverse fields of research. 
The impact of the specific problems we have discussed will differ by study purpose, which is why we consider the following a set of core questions to be considered by study authors---and not a definite guideline.

\textbf{Match Choice of List to Study Purpose:} Based on a precise understanding of what the domains in a list represent, an appropriate list type should be carefully chosen for a study. 
For example, the Umbrella list represents DNS names queried by many individual clients using OpenDNS (not only PCs, but also mobile devices and IoT devices), some bogus, some non-existent, but overall a representation of typical DNS traffic, and may form a good base for DNS analyses. 
The Alexa list gives a solid choice of functional websites frequently visited by users, and may be a good choice for a human web-centered study. 
Through its link-counting, the Majestic list also includes ``hidden'' links, and may include domains frequently loaded, but not necessarily knowingly requested by humans.
To obtain a reasonably general picture of the Internet, we recommend to scan a large sample, such as the ``general population'' used in \Cref{sec:impact}, \ie the set of all \textit{com/net/org} domains. 

\textbf{Consider Stability:} With lists changing up to 50\% per day, insights from measurement results might not even generalise to the next day.
For most measurement studies, stability should be increased by conducting repeated, longitudinal measurements.
This also helps to avoid bias from weekday vs.\ weekend lists.

\textbf{Document List and Measurement Details:} Studies should note the precise list (\eg Alexa Global \topm), its download date, and the measurements date to enable basic replicability. Ideally, the list used should be shared in a paper's dataset.

\subsection{Desired Properties for Top Lists}
Based on the challenges discussed in this work, we derive various properties that \toplists should offer:

\textbf{Consistency: }The characteristic, mainly structure and stability, of \toplists should be kept static over time. Where changes are required due to the evolving nature of the Internet, these should be announced and documented.

\textbf{Transparency: }Top list providers should be transparent about their ranking process and biases to help researchers understand and potentially control those biases. This may, of course, contradict the business interests of commercial list providers.

\textbf{Stability: }List stability faces a difficult trade-off: While capturing the ever-evolving trends in the Internet requires recent data, many typical \toplist uses are not stable to changes of up to 50\% per day. We hence suggest that lists should be offered as long-term (\eg a 90-day sliding window) and short-term (\eg only the most recent data) versions.

\subsection{Ethical Considerations}
We aim to minimise harm to all stakeholders possibly affected by our work. For active scans, we minimise interference by following best scanning practices~\cite{Durumeric2013}, such as maintaining a blacklist,  using dedicated servers with meaningful rDNS records, websites, and abuse contacts. 
We assess whether data collection can harm individuals and follow the beneficence principle as proposed  by~\cite{dittrich2012menlo,partridge2016ethical}.

Regarding list influencing in \Cref{sec:influencing}, the ethical implications of inserting a test domain into the \topm domains is small and unlikely to cause any harm.
In order to influence Umbrella ranks, we generated DNS traffic. For this, we selected query volumes unlikely to cause problems with the OpenDNS infrastructure or the RIPE Atlas platform.
Regarding the RIPE Atlas platform, we spread probes across the measurements as carefully as possible: 10k probes queried specific domains 100, 50, 10, and 1 times per day. In addition, 100, 1000, and 5000 probes performed an additional 100 queries per day. Per probe, that means 6,100 probes generated 261 queries per day (fewer than 11 queries per hour), and another 3,900 probes generated 161 queries per day. Refer to Figure~\ref{fig:umbrellahacking} to visualise the query volume.
That implies a total workload of around 2,220,000 queries per day. As the OpenDNS service is anycasted across multiple locations, it seems unlikely that our workload was a problem for the service.

 \section{Related Work}
We consider our work to be related to three fields:

\para{Sound Internet Measurements}
There exists a canon of work with guidelines on sound Internet measurements, such as \cite{paxson2004strategies, allmanculture,Durumeric2013,allman2017principles}.
These set out useful guidelines for measurements in general, but do not specifically tackle the issue of top lists.

\para{Measuring Web Popularity} 
Understanding web popularity is important for marketing as well as for business performance analyses. 
A book authored by Croll and Power~\cite{croll2009complete} warns site owners about the 
potential instrumentation biases present in Alexa ranks, specially with low-traffic sites.
Besides that, there is a set of blog posts and articles from the SEO space about anecdotal problems with certain top lists, but none of these conduct systematic analyses~\cite{alexabiasTechCruch,majesticahrefs}.

\para{Limitations of Using Top Lists in Research}
Despite the fact that top lists are widely used by research papers, we are not aware of any study focusing on the content of popular lists.
However, a number of research papers mentioned the limitations of relying on those ranks for their
specific research efforts~\cite{paper018:2017:ieeesp, paper029:2017:conext}.
W\"ahlisch \etal~\cite{wahlisch2015ripki} discuss the challenges of using top lists for web measurements.
They demonstrate that results vary when including \textit{www} subdomains, and investigate root causes such as routing failures. 
The aforementioned recent work by le\,Pochat \etal~\cite{pochat2018rigging} focuses on manipulating \toplists.

 \section{Conclusion}\label{sec:conclusions}
To the best of our knowledge, this is the first comprehensive study of the structure, stability, and significance of popular Internet \toplists.
We have shown that use of top lists is significant among networking papers, and found distinctive structural characteristics per list.
List stability has revealed interesting highlights, such as up to 50\% churn per day for some lists.
We have closely investigated ranking mechanisms of lists and manipulated a test domain's Umbrella rank in a controlled experiment.
Systematic measurement of \toplist domain characteristics and reproduction of studies has revealed that \toplists in general significantly distort results from the general population, and that results can depend on the day of week. 
We closed our work with a discussion on desirable properties of top lists and recommendations for top list use in science. 
We share code, data, and additional insights under
\begin{center}
	\url{https://toplists.github.io}
\end{center}
For long-term access, we provide an archival mirror at the TUM University Library: \url{https://mediatum.ub.tum.de/1452290}.

\textbf{Acknowledgements:} We thank the scientific community for the engaging discussions and data sharing leading to this publication, specifically Johanna Amann, Mark Allman, Matthias Wählisch, Ralph Holz, Georg Carle, Victor le Pochat, and the PAM'18 poster session participants. 
We thank the anonymous reviewers of the IMC'18 main and shadow PCs for their comments, and Zakir Durumeric for shepherding this work. 
This work was partially funded by the German Federal Ministry of Education and Research under project X-Check (grant 16KIS0530), by the DFG as part of the CRC 1053 MAKI, 
and the US National Science Foundation under grant number CNS-1564329.
\balance
\bibliographystyle{unsrt}

\begin{thebibliography}{100}

\bibitem{alexa}
{Alexa}.
\newblock {Top 1M sites}.
\newblock \url{https://www.alexa.com/topsites}, May 24, 2018.
\newblock
  \url{http://s3.dualstack.us-east-1.amazonaws.com/alexa-static/top-1m.csv.zip}.

\bibitem{umbrella}
{Cisco}.
\newblock {Umbrella Top 1M List}.
\newblock
  \url{https://umbrella.cisco.com/blog/blog/2016/12/14/cisco-umbrella-1-million/}.

\bibitem{majestic}
{Majestic}.
\newblock \url{https://majestic.com/reports/majestic-million/}, May 17, 2018.

\bibitem{majesticahrefs}
Matthew Woodward.
\newblock {Ahrefs vs Majestic SEO – 1 Million Reasons Why Ahrefs Is Better}.
\newblock
  \url{https://www.matthewwoodward.co.uk/experiments/ahrefs-majestic-seo-1-million-domain-showdown/},
  May 23, 2018.

\bibitem{alexatoolbarIE}
Alexa.
\newblock {The Alexa Extension}.
\newblock
  \url{https://web.archive.org/web/20160604100555/http://www.alexa.com/toolbar},
  June 04, 2016.

\bibitem{alexapanel}
{Alexa}.
\newblock {Alexa Increases its Global Traffic Panel}.
\newblock \url{https://blog.alexa.com/alexa-panel-increase/}, May 17, 2018.

\bibitem{alexamyths}
{Alexa}.
\newblock {Top 6 Myths about the Alexa Traffic Rank}.
\newblock
  \url{https://blog.alexa.com/top-6-myths-about-the-alexa-traffic-rank/}, May
  22, 2018.

\bibitem{alexalongtail}
{Alexa}.
\newblock {What’s going on with my Alexa Rank?}
\newblock \url{https://support.alexa.com/hc/en-us/articles/200449614}, May 17,
  2018.

\bibitem{majesticlaunch}
{Majestic}.
\newblock {Majestic Million CSV now free for all, daily}.
\newblock
  \url{https://blog.majestic.com/development/majestic-million-csv-daily/}, May
  17, 2018.

\bibitem{quantcast}
{Quantcast}.
\newblock \url{https://www.quantcast.com/top-sites/US/1}.

\bibitem{statvoo}
{Statvoo}.
\newblock \url{https://statvoo.com/top/sites}, May 17, 2018.

\bibitem{chromeuxr}
{Google}.
\newblock {Chrome User Experience Report}.
\newblock
  \url{https://developers.google.com/web/tools/chrome-user-experience-report/},
  May 15, 2018.

\bibitem{similarweb}
{SimilarWeb Top Websites Ranking}.
\newblock \url{https://www.similarweb.com/top-websites}.

\bibitem{paper003:2017:imc}
Vasileios Giotsas, Philipp Richter, Georgios Smaragdakis, Anja Feldmann,
  Christoph Dietzel, and Arthur Berger.
\newblock {Inferring BGP Blackholing Activity in the Internet}.
\newblock In {\em {Proceedings of the 2017 Internet Measurement Conference}},
  IMC '17, November 2017.

\bibitem{paper004:2017:imc}
Srikanth Sundaresan, Xiaohong Deng, Yun Feng, Danny Lee, and Amogh Dhamdhere.
\newblock {Challenges in Inferring Internet Congestion Using Throughput
  Measurements}.
\newblock In {\em {Proceedings of the 2017 Internet Measurement Conference}},
  IMC '17, November 2017.

\bibitem{paper013:2017:imc}
Zhongjie Wang, Yue Cao, Zhiyun Qian, Chengyu Song, and Srikanth~V.
  Krishnamurthy.
\newblock {Your State is Not Mine: A Closer Look at Evading Stateful Internet
  Censorship}.
\newblock In {\em {Proceedings of the 2017 Internet Measurement Conference}},
  IMC '17, November 2017.

\bibitem{paper016:2017:imc}
Savvas Zannettou, Tristan Caulfield, Emiliano De~Cristofaro, Nicolas
  Kourtelris, Ilias Leontiadis, Michael Sirivianos, Gianluca Stringhini, and
  Jeremy Blackburn.
\newblock {The Web Centipede: Understanding How Web Communities Influence Each
  Other Through the Lens of Mainstream and Alternative News Sources}.
\newblock In {\em {Proceedings of the 2017 Internet Measurement Conference}},
  IMC '17, November 2017.

\bibitem{paper028:2017:imc}
Austin Murdock, Frank Li, Paul Bramsen, Zakir Durumeric, and Vern Paxson.
\newblock {Target Generation for Internet-wide IPv6 Scanning}.
\newblock In {\em {Proceedings of the 2017 Internet Measurement Conference}},
  IMC '17, November 2017.

\bibitem{paper071:2017:imc}
Jan R\"{u}th, Christian Bormann, and Oliver Hohlfeld.
\newblock {Large-scale Scanning of TCP's Initial Window}.
\newblock In {\em {Proceedings of the 2017 Internet Measurement Conference}},
  IMC '17, November 2017.

\bibitem{paper123:2017:imc}
Umar Iqbal, Zubair Shafiq, and Zhiyun Qian.
\newblock {The Ad Wars: Retrospective Measurement and Analysis of Anti-adblock
  Filter Lists}.
\newblock In {\em {Proceedings of the 2017 Internet Measurement Conference}},
  IMC '17, November 2017.

\bibitem{paper173:2017:imc}
Johanna Amann, Oliver Gasser, Quirin Scheitle, Lexi Brent, Georg Carle, and
  Ralph Holz.
\newblock {Mission Accomplished?: HTTPS Security After Diginotar}.
\newblock In {\em {Proceedings of the 2017 Internet Measurement Conference}},
  IMC '17, November 2017.

\bibitem{paper176:2017:imc}
Joe DeBlasio, Stefan Savage, Geoffrey~M. Voelker, and Alex~C. Snoeren.
\newblock {Tripwire: Inferring Internet Site Compromise}.
\newblock In {\em {Proceedings of the 2017 Internet Measurement Conference}},
  IMC '17, November 2017.

\bibitem{paper177:2017:imc}
Shehroze Farooqi, Fareed Zaffar, Nektarios Leontiadis, and Zubair Shafiq.
\newblock {Measuring and Mitigating Oauth Access Token Abuse by Collusion
  Networks}.
\newblock In {\em {Proceedings of the 2017 Internet Measurement Conference}},
  IMC '17, November 2017.

\bibitem{paper219:2017:imc}
Janos Szurdi and Nicolas Christin.
\newblock {Email Typosquatting}.
\newblock In {\em {Proceedings of the 2017 Internet Measurement Conference}},
  IMC '17, November 2017.

\bibitem{paper061:2017:pam}
Enrico Bocchi, Luca De~Cicco, Marco Mellia, and Dario Rossi.
\newblock {The Web, the Users, and the MOS: Influence of HTTP/2 on User
  Experience}.
\newblock In {\em International Conference on Passive and Active Network
  Measurement}, pages 47--59. Springer, 2017.

\bibitem{paper063:2017:pam}
Ilker~Nadi Bozkurt, Anthony Aguirre, Balakrishnan Chandrasekaran, P~Brighten
  Godfrey, Gregory Laughlin, Bruce Maggs, and Ankit Singla.
\newblock {Why is the Internet so Slow?!}
\newblock In {\em International Conference on Passive and Active Network
  Measurement}, pages 173--187. Springer, 2017.

\bibitem{paper064:2017:pam}
Stephen Ludin.
\newblock {Measuring What is Not Ours: A Tale of 3rd Party Performance}.
\newblock In {\em Passive and Active Measurement: 18th International
  Conference, PAM 2017, Sydney, NSW, Australia, March 30-31, 2017,
  Proceedings}, volume 10176, page 142. Springer, 2017.

\bibitem{paper072:2017:pam}
Kittipat Apicharttrisorn, Ahmed Osama~Fathy Atya, Jiasi Chen, Karthikeyan
  Sundaresan, and Srikanth~V Krishnamurthy.
\newblock {Enhancing WiFi Throughput with PLC Extenders: A Measurement Study}.
\newblock In {\em International Conference on Passive and Active Network
  Measurement}, pages 257--269. Springer, 2017.

\bibitem{paper012:2017:tma}
Alexander Darer, Oliver Farnan, and Joss Wright.
\newblock {FilteredWeb: A framework for the Automated Search-based Discovery of
  Blocked URLs}.
\newblock In {\em Network Traffic Measurement and Analysis Conference (TMA),
  2017}, pages 1--9. IEEE, 2017.

\bibitem{paper027:2017:tma}
Jelena Mirkovic, Genevieve Bartlett, John Heidemann, Hao Shi, and Xiyue Deng.
\newblock {Do You See Me Now? Sparsity in Passive Observations of Address
  Liveness}.
\newblock In {\em Network Traffic Measurement and Analysis Conference (TMA),
  2017}, pages 1--9. IEEE, 2017.

\bibitem{paper107:2017:tma}
Quirin Scheitle, Oliver Gasser, Minoo Rouhi, and Georg Carle.
\newblock {Large-scale Classification of IPv6-IPv4 Siblings with Variable Clock
  Skew}.
\newblock In {\em Network Traffic Measurement and Analysis Conference (TMA),
  2017}, pages 1--9. IEEE, 2017.

\bibitem{paper007:2017:usenixsec}
Paul Pearce, Ben Jones, Frank Li, Roya Ensafi, Nick Feamster, Nick Weaver, and
  Vern Paxson.
\newblock {Global Measurement of DNS Manipulation}.
\newblock In {\em Proceedings of the 26th USENIX Security Symposium (USENIX
  Security '17)}, August 2017.

\bibitem{paper014:2017:usenixsec}
Rachee Singh, Rishab Nithyanand, Sadia Afroz, Paul Pearce, Michael~Carl
  Tschantz, Phillipa Gill, and Vern Paxson.
\newblock {Characterizing the Nature and Dynamics of Tor Exit Blocking}.
\newblock In {\em Proceedings of the 26th USENIX Security Symposium (USENIX
  Security '17)}, August 2017.

\bibitem{paper120:2017:usenixsec}
Tao Wang and Ian Goldberg.
\newblock {Walkie-Talkie: An Efficient Defense Against Passive Website
  Fingerprinting Attacks}.
\newblock In {\em Proceedings of the 26th USENIX Security Symposium (USENIX
  Security '17)}, August 2017.

\bibitem{paper122:2017:usenixsec}
Sebastian Zimmeck, Jie~S Li, Hyungtae Kim, Steven~M Bellovin, and Tony Jebara.
\newblock {A Privacy Analysis of Cross-device Tracking}.
\newblock In {\em Proceedings of the 26th USENIX Security Symposium (USENIX
  Security '17)}, August 2017.

\bibitem{paper146:2017:usenixsec}
Taejoong Chung, Roland van Rijswijk-Deij, Balakrishnan Chandrasekaran, David
  Choffnes, Dave Levin, Bruce~M Maggs, Alan Mislove, and Christo Wilson.
\newblock {A Longitudinal, End-to-End View of the DNSSEC Ecosystem}.
\newblock In {\em Proceedings of the 26th USENIX Security Symposium (USENIX
  Security '17)}, August 2017.

\bibitem{paper170:2017:usenixsec}
Katharina Krombholz, Wilfried Mayer, Martin Schmiedecker, and Edgar Weippl.
\newblock {"I Have No Idea What I’m Doing" – On the Usability of Deploying
  HTTPS}.
\newblock In {\em Proceedings of the 26th USENIX Security Symposium (USENIX
  Security '17)}, August 2017.

\bibitem{paper172:2017:usenixsec}
Adrienne~Porter Felt, Richard Barnes, April King, Chris Palmer, Chris Bentzel,
  and Parisa Tabriz.
\newblock {Measuring HTTPS Adoption on the Web}.
\newblock In {\em Proceedings of the 26th USENIX Security Symposium (USENIX
  Security '17)}, August 2017.

\bibitem{paper179:2017:usenixsec}
Ben Stock, Martin Johns, Marius Steffens, and Michael Backes.
\newblock {How the Web Tangled Itself:Uncovering the History of Client-Side Web
  (In)Security}.
\newblock In {\em Proceedings of the 26th USENIX Security Symposium (USENIX
  Security '17)}, August 2017.

\bibitem{paper181:2017:usenixsec}
Pepe Vila and Boris K{\"o}pf.
\newblock {Loophole: Timing Attacks on Shared Event Loops in Chrome}.
\newblock In {\em Proceedings of the 26th USENIX Security Symposium (USENIX
  Security '17)}, August 2017.

\bibitem{paper182:2017:usenixsec}
J{\"o}rg Schwenk, Marcus Niemietz, and Christian Mainka.
\newblock {Same-Origin Policy: Evaluation in Modern Browser}.
\newblock In {\em Proceedings of the 26th USENIX Security Symposium (USENIX
  Security '17)}, August 2017.

\bibitem{paper184:2017:usenixsec}
Stefano Calzavara, Alvise Rabitti, and Michele Bugliesi.
\newblock {CCSP: Controlled Relaxation of Content Security Policies by Runtime
  Policy Composition}.
\newblock In {\em Proceedings of the 26th USENIX Security Symposium (USENIX
  Security '17)}, August 2017.

\bibitem{paper232:2017:usenixsec}
Fang Liu, Chun Wang, Andres Pico, Danfeng Yao, and Gang Wang.
\newblock {Measuring the Insecurity of Mobile Deep Links of Android}.
\newblock In {\em Proceedings of the 26th USENIX Security Symposium (USENIX
  Security '17)}, August 2017.

\bibitem{paper011:2017:ieeesp}
Paul Pearce, Roya Ensafi, Frank Li, Nick Feamster, and Vern Paxson.
\newblock {Augur: Internet-Wide Detection of Connectivity Disruptions}.
\newblock In {\em IEEE Symposium on Security and Privacy}, 2017.

\bibitem{paper018:2017:ieeesp}
Sumayah Alrwais, Xiaojing Liao, Xianghang Mi, Peng Wang, Xiaofeng Wang, Feng
  Qian, Raheem Beyah, and Damon McCoy.
\newblock {Under the Shadow of Sunshine: Understanding and Detecting
  Bulletproof Hosting on Legitimate Service Provider Networks}.
\newblock In {\em IEEE Symposium on Security and Privacy}, 2017.

\bibitem{paper106:2017:ieeesp}
Oleksii Starov and Nick Nikiforakis.
\newblock {XHOUND: Quantifying the Fingerprintability of Browser Extensions}.
\newblock In {\em IEEE Symposium on Security and Privacy}, 2017.

\bibitem{paper144:2017:ieeesp}
Chaz Lever, Platon Kotzias, Davide Balzarotti, Juan Caballero, and Manos
  Antonakakis.
\newblock {A Lustrum of Malware Network Communication: Evolution and Insights}.
\newblock In {\em IEEE Symposium on Security and Privacy}, 2017.

\bibitem{paper208:2017:ieeesp}
James Larisch, David Choffnes, Dave Levin, Bruce~M. Maggs, Alan Mislove, and
  Christo Wilson.
\newblock {CRLite: A Scalable System for Pushing All TLS Revocations to All
  Browsers}.
\newblock In {\em IEEE Symposium on Security and Privacy}, 2017.

\bibitem{paper229:2017:ieeesp}
Nethanel Gelernter, Senia Kalma, Bar Magnezi, and Hen Porcilan.
\newblock {The Password Reset MitM Attack}.
\newblock In {\em IEEE Symposium on Security and Privacy}, 2017.

\bibitem{paper104:2017:ccs}
Milad Nasr, Amir Houmansadr, and Arya Mazumdar.
\newblock {Compressive Traffic Analysis: A New Paradigm for Scalable Traffic
  Analysis}.
\newblock In {\em CCS '17: Proceedings of the 2017 ACM SIGSAC Conference on
  Computer and Communications Security}, November 2017.

\bibitem{paper169:2017:ccs}
Daiping Liu, Zhou Li, Kun Du, Haining Wang, Baojun Liu, and Haixin Duan.
\newblock {Don't Let One Rotten Apple Spoil the Whole Barrel: Towards Automated
  Detection of Shadowed Domains}.
\newblock In {\em CCS '17: Proceedings of the 2017 ACM SIGSAC Conference on
  Computer and Communications Security}, November 2017.

\bibitem{paper174:2017:ccs}
Thomas Vissers, Timothy Barron, Tom Van~Goethem, Wouter Joosen, and Nick
  Nikiforakis.
\newblock {The Wolf of Name Street: Hijacking Domains Through Their
  Nameservers}.
\newblock In {\em CCS '17: Proceedings of the 2017 ACM SIGSAC Conference on
  Computer and Communications Security}, November 2017.

\bibitem{paper175:2017:ccs}
Ada Lerner, Tadayoshi Kohno, and Franziska Roesner.
\newblock {Rewriting History: Changing the Archived Web from the Present}.
\newblock In {\em CCS '17: Proceedings of the 2017 ACM SIGSAC Conference on
  Computer and Communications Security}, November 2017.

\bibitem{paper183:2017:ccs}
Yinzhi Cao, Zhanhao Chen, Song Li, and Shujiang Wu.
\newblock {Deterministic Browser}.
\newblock In {\em CCS '17: Proceedings of the 2017 ACM SIGSAC Conference on
  Computer and Communications Security}, November 2017.

\bibitem{paper185:2017:ccs}
Yizheng Chen, Yacin Nadji, Athanasios Kountouras, Fabian Monrose, Roberto
  Perdisci, Manos Antonakakis, and Nikolaos Vasiloglou.
\newblock {Practical Attacks Against Graph-based Clustering}.
\newblock In {\em CCS '17: Proceedings of the 2017 ACM SIGSAC Conference on
  Computer and Communications Security}, November 2017.

\bibitem{paper207:2017:ccs}
Sebastian Lekies, Krzysztof Kotowicz, Samuel Gro\ss, Eduardo~A. Vela~Nava, and
  Martin Johns.
\newblock {Code-Reuse Attacks for the Web: Breaking Cross-Site Scripting
  Mitigations via Script Gadgets}.
\newblock In {\em CCS '17: Proceedings of the 2017 ACM SIGSAC Conference on
  Computer and Communications Security}, November 2017.

\bibitem{paper216:2017:ccs}
Milad Nasr, Hadi Zolfaghari, and Amir Houmansadr.
\newblock {The Waterfall of Liberty: Decoy Routing Circumvention That Resists
  Routing Attacks}.
\newblock In {\em CCS '17: Proceedings of the 2017 ACM SIGSAC Conference on
  Computer and Communications Security}, November 2017.

\bibitem{paper220:2017:ccs}
Panagiotis Kintis, Najmeh Miramirkhani, Charles Lever, Yizheng Chen, Rosa
  Romero-G\'{o}mez, Nikolaos Pitropakis, Nick Nikiforakis, and Manos
  Antonakakis.
\newblock {Hiding in Plain Sight: A Longitudinal Study of Combosquatting
  Abuse}.
\newblock In {\em CCS '17: Proceedings of the 2017 ACM SIGSAC Conference on
  Computer and Communications Security}, November 2017.

\bibitem{paper230:2017:ccs}
Doowon Kim, Bum~Jun Kwon, and Tudor Dumitra\c{s}.
\newblock {Certified Malware: Measuring Breaches of Trust in the Windows
  Code-Signing PKI}.
\newblock In {\em CCS '17: Proceedings of the 2017 ACM SIGSAC Conference on
  Computer and Communications Security}, November 2017.

\bibitem{paper231:2017:ccs}
Peter Snyder, Cynthia Taylor, and Chris Kanich.
\newblock {Most Websites Don't Need to Vibrate: A Cost-Benefit Approach to
  Improving Browser Security}.
\newblock In {\em CCS '17: Proceedings of the 2017 ACM SIGSAC Conference on
  Computer and Communications Security}, November 2017.

\bibitem{paper121:2017:ndss}
Benjamin Greschbach, Tobias Pulls, Laura~M. Roberts, Phillip Winter, and Nick
  Feamster.
\newblock {The Effect of {DNS} on Tor's Anonymity}.
\newblock In {\em 24th Annual Network and Distributed System Security
  Symposium, {NDSS} 2017}, NDSS '17', February 2017.

\bibitem{paper206:2017:ndss}
Tobias Lauinger, Abdelberi Chaabane, Sajjad Arshad, William Robertson, Christo
  Wilson, and Engin Kirda.
\newblock {Thou Shalt Not Depend on Me: Analysing the Use of Outdated
  JavaScript Libraries on the Web}.
\newblock In {\em 24th Annual Network and Distributed System Security
  Symposium, {NDSS} 2017}, NDSS '17', February 2017.

\bibitem{paper217:2017:ndss}
Najmeh Miramirkhani, Oleksii Starov, and Nick Nikiforakis.
\newblock {Dial One for Scam: {A} Large-Scale Analysis of Technical Support
  Scams}.
\newblock In {\em 24th Annual Network and Distributed System Security
  Symposium, {NDSS} 2017}, NDSS '17', February 2017.

\bibitem{paper005:2017:conext}
Marc~Anthony Warrior, Uri Klarman, Marcel Flores, and Aleksandar Kuzmanovic.
\newblock {Drongo: Speeding Up CDNs with Subnet Assimilation from the Client}.
\newblock In {\em {CoNEXT '17: Proceedings of the 13th International Conference
  on Emerging Networking EXperiments and Technologies}}. ACM, December 2017.

\bibitem{paper008:2017:conext}
Shinyoung Cho, Rishab Nithyanand, Abbas Razaghpanah, and Phillipa Gill.
\newblock {A Churn for the Better}.
\newblock In {\em {CoNEXT '17: Proceedings of the 13th International Conference
  on Emerging Networking EXperiments and Technologies}}. ACM, December 2017.

\bibitem{paper015:2017:conext}
Wai~Kay Leong, Zixiao Wang, and Ben Leong.
\newblock {TCP Congestion Control Beyond Bandwidth-Delay Product for Mobile
  Cellular Networks}.
\newblock In {\em {CoNEXT '17: Proceedings of the 13th International Conference
  on Emerging Networking EXperiments and Technologies}}. ACM, December 2017.

\bibitem{paper029:2017:conext}
Mario Almeida, Alessandro Finamore, Diego Perino, Narseo Vallina-Rodriguez, and
  Matteo Varvello.
\newblock {Dissecting DNS Stakeholders in Mobile Networks}.
\newblock In {\em {CoNEXT '17: Proceedings of the 13th International Conference
  on Emerging Networking EXperiments and Technologies}}. ACM, December 2017.

\bibitem{paper145:2017:conext}
David Naylor, Richard Li, Christos Gkantsidis, Thomas Karagiannis, and Peter
  Steenkiste.
\newblock {And Then There Were More: Secure Communication for More Than Two
  Parties}.
\newblock In {\em {CoNEXT '17: Proceedings of the 13th International Conference
  on Emerging Networking EXperiments and Technologies}}. ACM, December 2017.

\bibitem{paper006:2017:sigcomm}
Thomas Holterbach, Stefano Vissicchio, Alberto Dainotti, and Laurent Vanbever.
\newblock Swift: Predictive fast reroute.
\newblock In {\em Proceedings of the Conference of the ACM Special Interest
  Group on Data Communication}, SIGCOMM '17. ACM, August 2017.

\bibitem{paper017:2017:sigcomm}
Costas Iordanou, Claudio Soriente, Michael Sirivianos, and Nikolaos Laoutaris.
\newblock {Who is Fiddling with Prices?: Building and Deploying a Watchdog
  Service for E-commerce}.
\newblock In {\em Proceedings of the Conference of the ACM Special Interest
  Group on Data Communication}, SIGCOMM '17. ACM, August 2017.

\bibitem{paper062:2017:sigcomm}
Vaspol Ruamviboonsuk, Ravi Netravali, Muhammed Uluyol, and Harsha~V.
  Madhyastha.
\newblock {Vroom: Accelerating the Mobile Web with Server-Aided Dependency
  Resolution}.
\newblock In {\em Proceedings of the Conference of the ACM Special Interest
  Group on Data Communication}, SIGCOMM '17. ACM, August 2017.

\bibitem{paper030:2017:www}
Sanae Rosen, Bo~Han, Shuai Hao, Z.~Morley Mao, and Feng Qian.
\newblock {Push or Request: An Investigation of HTTP/2 Server Push for
  Improving Mobile Web Performance}.
\newblock In {\em Proceedings of the 26th International Conference on World
  Wide Web}, 2017.

\bibitem{paper060:2017:www}
Elias~P. Papadopoulos, Michalis Diamantaris, Panagiotis Papadopoulos, Thanasis
  Petsas, Sotiris Ioannidis, and Evangelos~P. Markatos.
\newblock {The Long-Standing Privacy Debate: Mobile Websites vs Mobile Apps}.
\newblock In {\em Proceedings of the 26th International Conference on World
  Wide Web}, 2017.

\bibitem{paper105:2017:www}
Sanae Rosen, Bo~Han, Shuai Hao, Z.~Morley Mao, and Feng Qian.
\newblock {Extended Tracking Powers: Measuring the Privacy Diffusion Enabled by
  Browser Extensions}.
\newblock In {\em Proceedings of the 26th International Conference on World
  Wide Web}, 2017.

\bibitem{paper124:2017:www}
Deepak Kumar, Zane Ma, Zakir Durumeric, Ariana Mirian, Joshua Mason, J.~Alex
  Halderman, and Michael Bailey.
\newblock {Security Challenges in an Increasingly Tangled Web}.
\newblock In {\em Proceedings of the 26th International Conference on World
  Wide Web}, 2017.

\bibitem{paper134:2017:www}
Gareth Tyson, Shan Huang, Felix Cuadrado, Ignacio Castro, Vasile~C. Perta,
  Arjuna Sathiaseelan, and Steve Uhlig.
\newblock {Exploring HTTP Header Manipulation In-The-Wild}.
\newblock In {\em Proceedings of the 26th International Conference on World
  Wide Web}, 2017.

\bibitem{paper143:2017:www}
Luca Soldaini and Elad Yom-Tov.
\newblock {Inferring Individual Attributes from Search Engine Queries and
  Auxiliary Information}.
\newblock In {\em Proceedings of the 26th International Conference on World
  Wide Web}, 2017.

\bibitem{paper171:2017:www}
Kyungtae Kim, I~Luk Kim, Chung~Hwan Kim, Yonghwi Kwon, Yunhui Zheng, Xiangyu
  Zhang, and Dongyan Xu.
\newblock {J-Force: Forced Execution on JavaScript}.
\newblock In {\em Proceedings of the 26th International Conference on World
  Wide Web}, 2017.

\bibitem{paper204:2017:www}
Dolière~Francis Some, Nataliia Bielova, and Tamara Rezk.
\newblock {On the Content Security Policy Violations due to the Same-Origin
  Policy}.
\newblock In {\em Proceedings of the 26th International Conference on World
  Wide Web}, 2017.

\bibitem{paper205:2017:www}
Milivoj Simeonovski, Giancarlo Pellegrino, Christian Rossow, and Michael
  Backes.
\newblock {Who Controls the Internet? Analyzing Global Threats using Property
  Graph Traversals}.
\newblock In {\em Proceedings of the 26th International Conference on World
  Wide Web}, 2017.

\bibitem{paper209:2017:www}
Ajaya Neupane, Nitesh Saxena, and Leanne Hirshfield.
\newblock {Neural Underpinnings of Website Legitimacy and Familiarity
  Detection: An fNIRS Study}.
\newblock In {\em Proceedings of the 26th International Conference on World
  Wide Web}, 2017.

\bibitem{paper214:2017:www}
Qian Cui, Guy-Vincent Jourdan, Gregor Bochmann, Russell Couturier, and Vio
  Onut.
\newblock {Tracking Phishing Attacks Over Time}.
\newblock In {\em Proceedings of the 26th International Conference on World
  Wide Web}, 2017.

\bibitem{paper215:2017:www}
Li~Chang, Hsu-Chun Hsiao, Wei Jeng, Tiffany Hyun-Jin Kim, and Wei-Hsi Lin.
\newblock {Security Implications of Redirection Trail in Popular Websites
  Worldwide}.
\newblock In {\em Proceedings of the 26th International Conference on World
  Wide Web}, 2017.

\bibitem{paper218:2017:www}
Enrico Mariconti, Jeremiah Onaolapo, Sharique Ahmad, Nicolas Nikiforou, Manuel
  Egele, Nick Nikiforakis, and Gianluca Stringhini.
\newblock {What’s in a Name? Understanding Profile Name Reuse on Twitter}.
\newblock In {\em Proceedings of the 26th International Conference on World
  Wide Web}, 2017.

\bibitem{AcmArtifacts}
{ACM}.
\newblock {Result and Artifact Review and Badging}.
\newblock
  \url{https://www.acm.org/publications/policies/artifact-review-badging}, Acc.
  Jan 18 2017.

\bibitem{reproduc2017}
Quirin Scheitle, Matthias W{\"a}hlisch, Oliver Gasser, Thomas~C. Schmidt, and
  Georg Carle.
\newblock {Towards an Ecosystem for Reproducible Research in Computer
  Networking}.
\newblock In {\em ACM SIGCOMM 2017 Reproducibility Workshop}, 2017.

\bibitem{Flittner}
Matthias Flittner, Mohamed~Naoufal Mahfoudi, Damien Saucez, Matthias
  W\"{a}hlisch, Luigi Iannone, Vaibhav Bajpai, and Alex Afanasyev.
\newblock {A Survey on Artifacts from CoNEXT, ICN, IMC, and SIGCOMM Conferences
  in 2017}.
\newblock {\em SIGCOMM Comput. Commun. Rev.}, 48(1):75--80, April 2018.

\bibitem{saucez2018thoughts}
Damien Saucez and Luigi Iannone.
\newblock {Thoughts and Recommendations from the ACM SIGCOMM 2017
  Reproducibility Workshop}.
\newblock {\em ACM SIGCOMM Computer Communication Review}, 48(1):70--74, 2018.

\bibitem{allman2018robustness}
Mark Allman.
\newblock {Comments On DNS Robustness}.
\newblock {\em IMC}, 2018.

\bibitem{wahlisch2015ripki}
Matthias W{\"a}hlisch, Robert Schmidt, Thomas~C Schmidt, Olaf Maennel, Steve
  Uhlig, and Gareth Tyson.
\newblock {RiPKI: The tragic story of RPKI deployment in the Web ecosystem}.
\newblock In {\em Proceedings of the 14th ACM Workshop on Hot Topics in
  Networks}, page~11. ACM, 2015.

\bibitem{holzimc2011}
Ralph Holz, Lothar Braun, Nils Kammenhuber, and Georg Carle.
\newblock {The SSL Landscape - A Thorough Analysis of the X.509 PKI Using
  Active and Passive Measurements}.
\newblock In {\em IMC}, Nov. 2011.

\bibitem{archivealexacrawl}
{The Internet Archive}.
\newblock {Alexa Crawls}.
\newblock \url{https://archive.org/details/alexacrawls}, May 22, 2018.

\bibitem{pslgithub}
Mozilla.
\newblock {Public Suffix List}: commit 2f9350.
\newblock \url{https://github.com/publicsuffix/list/commit/85fa8fbdf}, Apr. 20,
  2018.

\bibitem{ianatld}
{IANA}.
\newblock {TLD Directory}.
\newblock http://data.iana.org/TLD/tlds-alpha-by-domain.txt, May 20, 2018.

\bibitem{terminatedtlds}
ICANN.
\newblock {Notices of Termination and Status of gTLD}.
\newblock
  \url{https://www.icann.org/resources/pages/gtld-registry-agreement-termination-2015-10-09-en},
  Apr. 20, 2018.

\bibitem{stopusingio}
Nick Parsons.
\newblock {Stop using .IO Domain Names for Production Traffic}.
\newblock
  \url{https://hackernoon.com/stop-using-io-domain-names-for-production-traffic-b6aa17eeac20},
  May 21, 2018.

\bibitem{ioerror}
Matthew Bryant.
\newblock {The .io Error – Taking Control of All .io Domains With a Targeted
  Registration}.
\newblock
  \url{https://thehackerblog.com/the-io-error-taking-control-of-all-io-domains-with-a-targeted-registration/},
  May 21, 2018.

\bibitem{ioregistrar}
Tomislav Lombarovic.
\newblock {Be aware: How domain registrar can kill your business}.
\newblock
  \url{https://www.uptimechecker.io/blog/how-domain-registrar-can-kill-your-business},
  May 21, 2018.

\bibitem{newtldtimelines}
ICANN.
\newblock {new gTLD Program Timeline}.
\newblock \url{https://newgtlds.icann.org/en/program-status/timelinesen}, Apr.
  20, 2018.

\bibitem{razaghpanah2018apps}
Abbas Razaghpanah, Rishab Nithyanand, Narseo Vallina-Rodriguez, Srikanth
  Sundaresan, Mark Allman, Christian Kreibich, and Phillipa Gill.
\newblock {Apps, Trackers, Privacy, and Regulators: A Global Study of the
  Mobile Tracking Ecosystem}.
\newblock In {\em NDSS}, 2018.

\bibitem{opendnsdomaintagging}
OpenDNS.
\newblock {Domain Tagging}.
\newblock \url{https://domain.opendns.com}.

\bibitem{razaghpanah2015haystack}
Abbas Razaghpanah, Narseo Vallina-Rodriguez, Srikanth Sundaresan, Christian
  Kreibich, Phillipa Gill, Mark Allman, and Vern Paxson.
\newblock {Haystack: A multi-purpose mobile vantage point in user space}.
\newblock {\em arXiv preprint arXiv:1510.01419}, 2015.

\bibitem{hphosts}
{hpHosts}.
\newblock {hpHosts Domain Blacklist}, May 21, 2018.
\newblock \url{https://hosts-file.net/}.

\bibitem{lakhina2004structural}
Anukool Lakhina, Konstantina Papagiannaki, Mark Crovella, Christophe Diot,
  Eric~D Kolaczyk, and Nina Taft.
\newblock {Structural Analysis of Network Traffic Flows}.
\newblock In {\em {ACM SIGMETRICS Performance Evaluation Review}}, volume~32,
  pages 61--72. ACM, 2004.

\bibitem{papagiannaki2003long}
Konstantina Papagiannaki, Nina Taft, Z-L Zhang, and Christophe Diot.
\newblock {Long-term Forecasting of Internet Backbone Traffic: Observations and
  Initial Models}.
\newblock In {\em INFOCOM 2003. Twenty-Second Annual Joint Conference of the
  IEEE Computer and Communications. IEEE Societies}, volume~2, pages
  1178--1188. IEEE, 2003.

\bibitem{gill2007youtube}
Phillipa Gill, Martin Arlitt, Zongpeng Li, and Anirban Mahanti.
\newblock {Youtube Traffic Characterization: A View from the Edge}.
\newblock In {\em Proceedings of the 7th ACM SIGCOMM conference on Internet
  measurement}, pages 15--28. ACM, 2007.

\bibitem{cha2007tube}
Meeyoung Cha, Haewoon Kwak, Pablo Rodriguez, Yong-Yeol Ahn, and Sue Moon.
\newblock {I Tube, You Tube, Everybody Tubes: Analyzing the World's Largest
  User Generated Content Video System}.
\newblock In {\em Proceedings of the 7th ACM SIGCOMM conference on Internet
  measurement}, pages 1--14. ACM, 2007.

\bibitem{kendall1938new}
Maurice~G Kendall.
\newblock {A New Measure of Rank Correlation}.
\newblock {\em Biometrika}, 30(1/2):81--93, 1938.

\bibitem{alexahowranking}
{Alexa}.
\newblock {How are Alexa’s traffic rankings determined?}
\newblock \url{https://support.alexa.com/hc/en-us/articles/200449744}, May 17,
  2018.

\bibitem{alexaacquired}
Adam Feuerstein.
\newblock {E-commerce loves Street: Critical Path plans encore}.
\newblock San Francisco Business Times,
  \url{https://www.bizjournals.com/sanfrancisco/stories/1999/05/24/newscolumn4.html},
  May 1999.

\bibitem{alexaboostspecialist}
{Alexa Specialist}.
\newblock \url{http://www.improvealexaranking.com/}, May 22, 2018.

\bibitem{alexaboostrankboostup}
{Rankboostup}.
\newblock \url{https://rankboostup.com/}, May 22, 2018.

\bibitem{alexaboostupmyrank}
{UpMyRank}.
\newblock \url{http://www.upmyrank.com/}, May 22, 2018.

\bibitem{pochat2018rigging}
Victor~Le Pochat, Tom Van~Goethem, and Wouter Joosen.
\newblock {Rigging Research Results by Manipulating Top Websites Rankings}.
\newblock {\em arXiv preprint arXiv:1806.01156}, June 4, 2018.

\bibitem{RA1}
{RIPE Atlas}.
\newblock {Measurement IDs 124307\{26,28-33\}}.

\bibitem{RA2}
{RIPE Atlas}.
\newblock {Measurement IDs 124674\{03-10\}}.

\bibitem{majesticabout}
{Majestic}.
\newblock {About Majestic}.
\newblock \url{https://blog.majestic.com/about/}, May 22, 2018.

\bibitem{majesticnewalgorithm}
{Majestic}.
\newblock {Majestic Million – Reloaded!}
\newblock \url{https://blog.majestic.com/company/majestic-million-reloaded/},
  May 22, 2018.

\bibitem{majesticinsights}
{Majestic}.
\newblock {A Million here… A Million there…}.
\newblock
  \url{https://blog.majestic.com/case-studies/a-million-here-a-million-there/},
  May 22, 2018.

\bibitem{pagerank}
Lawrence Page, Sergey Brin, Rajeev Motwani, and Terry Winograd.
\newblock The pagerank citation ranking: Bringing order to the web.
\newblock Technical report, Stanford InfoLab, 1999.

\bibitem{dni}
Verisign.
\newblock {The Domain Name Industry Brief 2017Q4}, 2018.

\bibitem{caastudy17}
Quirin Scheitle, Taejoong Chung, Jens Hiller, Oliver Gasser, Johannes Naab,
  Roland van Rijswijk-Deij, Oliver Hohlfeld, Ralph Holz, Dave Choffnes, Alan
  Mislove, and Georg Carle.
\newblock {A First Look at Certification Authority Authorization (CAA)}.
\newblock {\em ACM SIGCOMM CCR}, April 2018.

\bibitem{openintel_jsac}
Roland van Rijswijk{-}Deij, Mattijs Jonker, Anna Sperotto, and Aiko Pras.
\newblock {A High-Performance, Scalable Infrastructure for Large-Scale Active
  DNS Measurements}.
\newblock {\em {IEEE} {JSAC}}, 2016.

\bibitem{ipv6hitlist}
Oliver Gasser, Quirin Scheitle, Sebastian Gebhard, and Georg Carle.
\newblock {Scanning the IPv6 Internet: Towards a Comprehensive Hitlist}.
\newblock In {\em {TMA}}, 2016.

\bibitem{h2adoption17}
T.~Zimmermann, J.~R{\"u}th, B.~Wolters, and O.~Hohlfeld.
\newblock {How HTTP/2 pushes the web: An empirical study of HTTP/2 Server
  Push}.
\newblock In {\em 2017 IFIP Networking Conference (IFIP Networking) and
  Workshops}, pages 1--9, June 2017.

\bibitem{czyzv6}
Jakub Czyz, Mark Allman, Jing Zhang, Scott Iekel-Johnson, Eric Osterweil, and
  Michael Bailey.
\newblock {Measuring IPv6 Adoption}.
\newblock In {\em {ACM SIGCOMM}}, 2014.

\bibitem{eravuchira2016measuring}
Steffie~Jacob Eravuchira, Vaibhav Bajpai, J{\"u}rgen Sch{\"o}nw{\"a}lder, and
  Sam Crawford.
\newblock {Measuring web similarity from dual-stacked hosts}.
\newblock In {\em Network and Service Management (CNSM), 2016 12th
  International Conference on}, pages 181--187. IEEE, 2016.

\bibitem{ruohonen2018empirical}
Jukka Ruohonen.
\newblock {An Empirical Survey on the Early Adoption of DNS Certification
  Authority Authorization}.
\newblock {\em arXiv preprint arXiv:1804.07604}, 2018.

\bibitem{wptcdndns}
Google.
\newblock {WebPagetest CDN domain list, cdn.h}.
\newblock
  \url{https://github.com/WPO-Foundation/webpagetest/blob/master/agent/wpthook/cdn.h}.

\bibitem{routeviews}
University of~Oregon.
\newblock {Route Views Project}.
\newblock \url{http://www.routeviews.org}.

\bibitem{webh2yet16}
Matteo Varvello, Kyle Schomp, David Naylor, Jeremy Blackburn, Alessandro
  Finamore, and Konstantina Papagiannaki.
\newblock {Is the Web HTTP/2 Yet?}
\newblock In Thomas Karagiannis and Xenofontas Dimitropoulos, editors, {\em
  Passive and Active Measurement}, pages 218--232, Cham, 2016. Springer
  International Publishing.

\bibitem{Durumeric2013}
Zakir Durumeric, Eric Wustrow, and J.~Alex Halderman.
\newblock {ZMap: Fast {Internet}-wide Scanning and Its Security Applications}.
\newblock In {\em {USENIX} Security}, 2013.

\bibitem{dittrich2012menlo}
David Dittrich, Erin Kenneally, et~al.
\newblock {The Menlo Report: Ethical Principles Guiding Information and
  Communication Technology Research}.
\newblock {\em US Department of Homeland Security}, 2012.

\bibitem{partridge2016ethical}
Craig Partridge and Mark Allman.
\newblock {Ethical Considerations in Network Measurement Papers}.
\newblock {\em Communications of the ACM}, 2016.

\bibitem{paxson2004strategies}
Vern Paxson.
\newblock {Strategies for Sound Internet Measurement}.
\newblock In {\em Proceedings of the 4th ACM SIGCOMM conference on Internet
  measurement}, pages 263--271. ACM, 2004.

\bibitem{allmanculture}
Mark Allman.
\newblock {{On Changing the Culture of Empirical Internet Assessment}}.
\newblock {\em ACM Computer Communication Review}, 43(3), July 2013.
\newblock Editorial Contribution.

\bibitem{allman2017principles}
Mark Allman, Robert Beverly, and Brian Trammell.
\newblock Principles for measurability in protocol design.
\newblock {\em ACM SIGCOMM Computer Communication Review}, 47(2):2--12, 2017.

\bibitem{croll2009complete}
Alistair Croll and Sean Power.
\newblock {\em {Complete web monitoring: watching your visitors, performance,
  communities, and competitors}}.
\newblock " O'Reilly Media, Inc.", 2009.

\bibitem{alexabiasTechCruch}
{Michael Arrington}.
\newblock {Alexa’s Make Believe Internet; Alexa Says YouTube Is Now Bigger
  Than Google. Alexa Is Useless}.
\newblock
  \url{https://techcrunch.com/2007/11/25/alexas-make-believe-internet/}, 2007.

\end{thebibliography}


{\small

 }
\end{document}